\documentclass[sigconf]{acmart}

\usepackage{multirow}
\usepackage{color}
\usepackage{colortbl}
\usepackage{stfloats}
\usepackage{longfbox} 
\usepackage{subcaption}
\usepackage{enumitem}
\usepackage{graphicx}
\usepackage{subcaption}
\usepackage{tabularx}
\usepackage{longtable}

\makeatletter
\newdimen\@tempdimd
\makeatother

\definecolor{quotebackground}{HTML}{EFEFEF}
\definecolor{tableheader}{HTML}{EFEFEF}
\definecolor{tablegrayline}{HTML}{e0e0e0}

\newcommand{\eg}{\textit{e.g.}}
\newcommand{\ie}{\textit{i.e.}}
\newcommand{\cf}{\textit{c.f.}}
\newcommand{\etal}{\textit{et al.}}

\newcommand{\labelphantom}[1]{%
  \parbox{0pt}{\phantomsubcaption\label{#1}}%
}

\newcommand{\circledigit}[1]{\textbf{\normalsize{\textsf{\textcircled{\footnotesize{#1}}}}}}

\newcommand{\ipstart}[1]{\vspace{1mm} \noindent{\textbf{\textit{#1.}}}}

\newcommand{\npstart}[1]{\vspace{1mm} \noindent{{#1}}}

\newfboxstyle{scenarioparam}{padding=8pt, margin-top=3pt, margin-bottom=3pt, border-style=none, background-color=quotebackground}

\definecolor{minimalcolor}{HTML}{D3E593}
\definecolor{mildcolor}{HTML}{F7C38B}
\definecolor{severecolor}{HTML}{F2A19C}
\newfboxstyle{patternparam}{padding=2pt, border-style=none, height=6pt, border-radius=2pt}
\newcommand{\cesminimalbox}[1]{\lfbox[patternparam, background-color=minimalcolor]{#1}}
\newcommand{\cesmildbox}[1]{\lfbox[patternparam, background-color=mildcolor]{#1}}
\newcommand{\cesseverebox}[1]{\lfbox[patternparam, background-color=severecolor]{#1}}

\usepackage{newfloat}
\DeclareFloatingEnvironment[fileext=fml,placement={H},name=Dialogue]{dialogue}
\newenvironment{quotetable}{
\vspace{3mm}
\hfill\break
\noindent
    \small\sffamily\centering
    \def\arraystretch{1.2}\setlength{\tabcolsep}{0.25em}
    \tabularx{\columnwidth}{p{0.06\columnwidth}>{\raggedright\let\newline\\\arraybackslash\hspace{0pt}}m{0.9\columnwidth}}
    \hline}
{
    \arrayrulecolor{black}\hline
    \endtabularx
    \vspace{3mm}
}

\newcommand{\quotebotline}[1]{\textbf{AI} & \textit{#1}\\\arrayrulecolor{tablegrayline}\hline}

\newcommand{\quoteuserline}[2]{\textbf{#1} & 
\textit{#2} \\\arrayrulecolor{tablegrayline}\hline}

\AtBeginDocument{%
  \providecommand\BibTeX{{%
    \normalfont B\kern-0.5em{\scshape i\kern-0.25em b}\kern-0.8em\TeX}}}

\copyrightyear{2024}
\acmYear{2024}
\setcopyright{rightsretained}
\acmConference[CHI '24]{Proceedings of the CHI Conference on Human Factors in Computing Systems}{May 11--16, 2024}{Honolulu, HI, USA}
\acmBooktitle{Proceedings of the CHI Conference on Human Factors in Computing Systems (CHI '24), May 11--16, 2024, Honolulu, HI, USA}
\acmDOI{10.1145/3613904.3642937}
\acmISBN{979-8-4007-0330-0/24/05}

\usepackage{graphicx}
\usepackage{array}
\usepackage{kotex}

\acmSubmissionID{3044}

\newcommand{\red}[1]{#1}

\newcommand{\changed}[1]{\textcolor{black}{#1}}

\begin{document}

\title{MindfulDiary: Harnessing Large Language Model to~Support~Psychiatric Patients' Journaling}

\settopmatter{authorsperrow=4}
\author{Taewan Kim}
\authornote{Taewan Kim conducted this work as a research intern at NAVER AI Lab.}
\orcid{0000-0001-8578-5342}
\affiliation{%
  \institution{KAIST}
  \country{Republic of Korea}
}
\email{taewan@kaist.ac.kr}

\author{Seolyeong Bae}
\authornote{Seolyeong Bae conducted this work as an engineering intern at NAVER Cloud.}
\affiliation{%
  \institution{GIST}
  \country{Republic of Korea}
}
\email{peixueying@gmail.com}

\author{Hyun Ah Kim}
\affiliation{%
  \institution{NAVER Cloud}
  \country{Republic of Korea}
}
\email{hyunah.kim@navercorp.com}

\author{Su-woo Lee}
\affiliation{%
  \institution{Wonkwang Univ. Hospital}
  \country{Republic of Korea}
}
\email{aiesw@naver.com}

\author{Hwajung Hong}
\orcid{0000-0001-5268-3331}
\affiliation{%
  \institution{KAIST}
  \country{Republic of Korea}
}
\email{hwajung@kaist.ac.kr}

\author{Chanmo Yang}
\authornote{Co-corresponding authors.}
\affiliation{%
  \institution{Wonkwang Univ. Hospital, \newline{}Wonkwang University}
  \country{Republic of Korea}
}
\email{ychanmo@wku.ac.kr}

\author{Young-Ho Kim}
\authornotemark[3]
\orcid{0000-0002-2681-2774}
\affiliation{%
  \institution{NAVER AI Lab}
  \country{Republic of Korea}
}
\email{yghokim@younghokim.net}


\begin{abstract}
    Large Language Models (LLMs) offer promising opportunities in mental health domains, although their inherent complexity and low controllability elicit concern regarding their applicability in clinical settings. We present MindfulDiary, an LLM-driven journaling app that helps psychiatric patients document daily experiences through conversation. Designed in collaboration with mental health professionals, MindfulDiary takes a state-based approach to safely comply with the experts' guidelines while carrying on free-form conversations. Through a four-week field study involving 28 patients with major depressive disorder and five psychiatrists, we examined how MindfulDiary facilitates patients' journaling practice and clinical care. The study revealed that MindfulDiary supported patients in consistently enriching their daily records and helped clinicians better empathize with their patients through an understanding of their thoughts and daily contexts. Drawing on these findings, we discuss the implications of leveraging LLMs in the mental health domain, bridging the technical feasibility and their integration into clinical settings.
\end{abstract}



\begin{teaserfigure}
    \includegraphics[width=\textwidth]{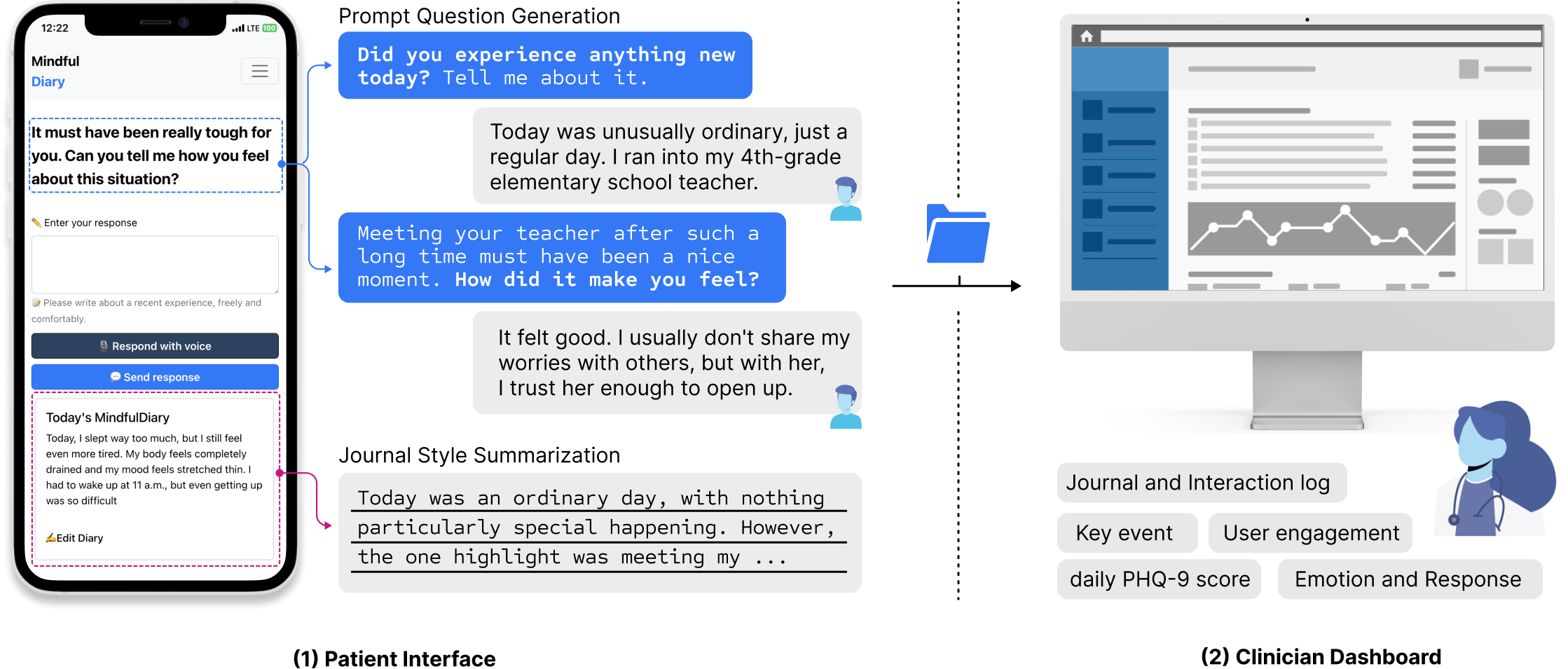}
    \caption{MindfulDiary consists of two components: the patient interface and the expert interface: (1) Patient interface aids users in daily journaling, offering prompt questions through conversations and crafting summaries in a diary-like format. (2) Clinician Dashboard features a dashboard that visualizes data from patient records, facilitating its use during consultations. (Refer to our supplementary video for the user interaction.)}
    \label{fig:teaser}
    \Description{Image of MindfulDiary's interface split into two sections. On the left, the patient interface displays a journal session screen which aids user reflection through prompt question generation and journal-style summarization. On the right, the expert interface showcases a desktop dashboard layout, providing professionals with insights into journal interactions, key events and emotions, as well as the trend of PHQ-9 scores.}
\end{teaserfigure}

\maketitle

    \section{Introduction}
Journals serve as a written record of an individual's past events, thoughts, and feelings, allowing genuine expression~\cite{Travers2011, Ullrich2002}. Journaling helps people describe experiences and express emotions related to both negative~\cite{pennebaker1997writing, pennebaker2011expressive} and positive experiences (\eg, growth potential)~\cite{Baikie2012, Guastella2009}, thereby reducing stress, anxiety, and depression. Prior work has shown the advantages of journaling in clinical mental health contexts, as journals frequently capture patients' daily experiences, symptoms, and other contextual data that are challenging to gather during brief hospital visits~\cite{figueiredo2020patient, zhu2016sharing}. Furthermore, these patient journals can enhance mental health professionals (MHPs) comprehension of their patient's conditions, leading to improved treatment quality~\cite{wu2020clinician}. However, writing about one's past feelings and thoughts can be a complex process because people differ in their ability to understand, identify, and verbalize their emotions~\cite{salovey1990}. In addition, patient under psychotherapy struggle with constructing a narrative and understanding their past~\cite{donnelly1991cognitive, pennebaker1999forming}.

Conversational AIs, or chatbots, have the potential as an alternative form of journaling, easing the collection of personal data. Researchers in the field of Human-Computer Interaction (HCI) have shown that chatbots can help individuals articulate and share their daily experiences. For instance, chatbots to elicit people’s self-disclosure can ease the process of emotional expression by providing a safe and supportive environment for individuals to share their experiences and emotions~\cite{park_chatbot, 10.1145/3313831.3376175, 10.1145/3392836, cameron2019assessing}. Furthermore, a machine's inherent trait of not showing fatigue can make people more confident to share their stories truthfully and comfortably~\cite{10.1145/3170427.3188548, park_chatbot}. However, existing chatbot prototypes have commonly employed rule-based or retrieval-driven approaches~\cite{ABDALRAZAQ2019103978}, which have limited capability of generating versatile responses following up serendipitous topics during conversation~\cite{Huang2020ChallengesODD, lee_caring_2019, Jo2023CareCallInterviews}. \changed{This trend presents missed opportunities and a lack of understanding regarding conversational AIs that assist with journaling by suggesting, questioning, and empathizing based on the user's diverse experiences.}

\red{The recent achievement of Natural Language Processing in large language models (LLMs) opened up new opportunities for bootstrapping chatbots that can carry on more naturalistic conversation~\cite{bae2022building, Jo2023CareCallInterviews, wei2023leveraging, roller-etal-2021-recipes, brown2020language}. Their capabilities accelerated the development of chatbots in varied topics that can benefit from open-ended conversation, such as regular check-up calls~\cite{Jo2023CareCallInterviews, bae2022building}, personal health tracking~\cite{wei2023leveraging}, and personal events and emotions~\cite{seo2023chacha}. Despite such opportunities, LLMs' inherent uncertainty in control of response generation calls for precautions to handle unintended or inaccurate responses~\cite{farhat2023chatgpt, chatbot_review21, npj21, Jo2023CareCallInterviews}. If applied to clinical and mental health domains, LLM's behaviors should be designed in collaboration with domain experts regarding the relevance and safety of responses.}

\red{In this work, we present a case of collaborative design, development, and evaluation of an LLM-infused conversational AI system designed to facilitate the self-reflection of patients and communication with MHPs. We designed and developed \textbf{MindfulDiary} (\autoref{fig:teaser}), which consists of (1) a mobile conversational AI with which patients can converse about daily experiences and thoughts and (2) a web dashboard that allows MHPs to review their patients' dialogue history with the AI.} MindfulDiary incorporates LLMs to generate a response, prompting patients differently according to the conversational phase. The conversation records are automatically summarized and presented on a clinician dashboard so MHPs can obtain insights about the patient.

As a multi-disciplinary research team, which included HCI researchers, AI engineers, and psychiatrists, we iteratively designed MindfulDiary and conducted a four-week field study involving 28 psychiatric patients diagnosed with major depressive disorder (MDD) and five psychiatrists who care for them. \red{During the study, the patients freely used MindfulDiary to record daily conversations, and the psychiatrists used the clinician dashboard during regular clinical visits.} Through this study, we found that the versatility, narrative-building capability, and diverse perspectives provided by MindfulDiary assisted patients in consistently enriching their daily records. Furthermore, MindfulDiary supported patients in overcoming the challenges of detailed record-keeping and expression, often hindered by feelings of apathy and cognitive burdens. \red{The psychiatrists reported that enhanced records provided by MindfulDiary offered a more nuanced understanding of their patients, fostering empathy. In addition, MindfulDiary supplemented their consultation by eliciting candid thoughts from patients that may be invasive to be asked by the MHPs.}

\npstart{The key contributions of this work are: }
\begin{enumerate}[leftmargin=*, itemsep=4pt, topsep=0pt]
    \item Design and development of MindfulDiary, an LLM-driven journal designed to document psychiatric patients' daily experiences through naturalistic conversations, designed in collaboration with MHPs.
    \item Empirical findings from a four-week field study involving 28 patients and five psychiatrists, demonstrating how MindfulDiary supported patients in keeping their daily logs and assisted psychiatrists in monitoring and comprehending patient states. \red{We also explore how MindfulDiary enhances the quality of patient-provider communication, emphasizing the role of LLMs in prompting deeper self-exploration, which can be instrumental in clinical settings.}
    \item \red{Implications for designing and instrumenting LLM-infused conversational AIs in clinical mental health settings.}
\end{enumerate}

    \section{Related Work}
In this section, we cover related work in three parts: (1) Journaling for self-reflection and mental Health, (2) Journaling as patient-generated health data in clinical settings, and (3) Conversational agents for mental health.

\subsection{Journaling for Self-Reflection and Mental Health}
Journaling---recording various personal matters ranging from observations and travels to overall daily experiences and thoughts~\cite{hategan2020humanism}---takes place in various forms, including letter-style entries reminiscent of Anne Frank's diary or more traditional prose. By narrating personal experiences and innermost feelings in a journal, people discover insights upon reflecting on their past events~\cite{Ullrich2002, Travers2011}, thereby improving mental wellness and quality of life~\cite{pennebaker1997writing, pennebaker2011expressive, Baikie2012, Utley2011}. Researchers have found that journaling impacts mental health by fostering emotional inhibition~\cite{pennebaker1985traumatic}, cognitive processing~\cite{harber1992overcoming}, and “freeing up” cognitive load~\cite{Klein2001, ramirez2011writing}.

\changed{Beyond traditional pen and paper, research in HCI has explored how technology can augment a journaling method in articulating past emotions and experiences~(\eg,~\cite{10.1145/2858036.2858103, kalnikaite2012synergetic, hodges2006sensecam, hodges2006sensecam, 10.1145/3214273, bakker2018engagement, Apple2024iPhone}). Early studies in journaling focused on how technology can assist users in better documenting past events and experiences from life-logging perspectives~\cite{10.1145/2858036.2858103, kalnikaite2012synergetic, hodges2006sensecam}. For example, SenseCam has been proposed as a wearable ubiquitous computing device that utilizes camera sensor data to not only record but also reflect on the wearer's daily life~\cite{hodges2006sensecam}. Subsequently, various attempts have been made to utilize contextual data and cues such as activity levels~\cite{10.1145/3214273}, mood~\cite{bakker2018engagement}, location, and photos~\cite{Apple2024iPhone} in journaling context.}

\changed{Furthermore, HCI researchers have investigated ways to improve a journaling method to encourage users to express themselves in a manner that is more comfortable~\cite{gonzales2010}, engaging~\cite{ma2017}, and honest~\cite{park_chatbot}.} For example, Park~\etal{} found that conversational agents create a social-like environment, encouraging self-reflection and enhancing expressive writing~\cite{park_chatbot}. Furthermore, social support from online communities has been found to increase user participation in journaling activities~\cite{ma2017}. Gonzales~\etal~\cite{gonzales2010} proposed an approach to mitigate the discomfort of revisiting negative memories using the sound generated from the data for more pleasurable~\cite{gonzales2010}.

Our work extends this line of research to enhance the journaling experience with technology. \changed{Specifically, we explore the potential of interactive dialogue with LLMs as a new format for journaling, allowing psychiatric patients to explore and reflect on their past experiences and emotions.} To this end, we leverage a conversational agent as a complementary tool for assisting patients by providing prompt questions to engage users in deeper and more detailed documentation and reflection~\cite{10.1145/3214273}. We particularly investigate how our approach plays a role in clinical settings as patient-generated health data, which we will cover in the following.

\subsection{Journaling as Patient-Generated Health Data in Clinical Settings}
\red{Patient-Generated Health Data (PGHD)}---defined as \textit{``health-related data, such as health history, symptoms, biometric readings, treatment history, lifestyle choices, and other pertinent details, created, recorded, or gathered by patients''}~\cite{shapiro2012patient}---has increasingly become an essential tool in clinical settings to capture authentic, real-time insights into patients` health. Studies have shown that PGHD can enhance communication between patients and MHPs and offer contextual information about patients, thereby heightening MHPs' awareness of patient health outside regular clinical visits~\cite{10.1177/1460458220928184, cohen2016integrating, nundy2014using}. \red{For instance, photo journaling improved patient-provider communication for the management and treatment of irritable bowel syndrome by facilitating more effective discussions during treatment~\cite{10.1145/3123024.3123197}.}

Within the mental health domain, PGHD range from structured mental health assessments (\eg, anxiety, depression) to more unstructured data, including mood-related symptoms and social interactions tracking (\eg, social media use, number of calls)\cite{austin2020use, demiris2019patient, nghiem2023understanding}. Patient diaries or journals--\textit{``instructing the patient to write down one's symptoms and other information related to one's daily life to discuss them during clinical appointments''}~\cite{figueiredo2020patient, zhu2016sharing}--can particularly be useful in mental health contexts as it can offer rich, self-documented insights, which could improve MHPs` understanding of their patients~\cite{wu2020clinician}. Despite the benefits of patient journaling, people often struggle with starting their entries, sticking to consistent journaling routines, and structuring their reflections~\cite{Travers2011}. Further, writing about emotions and past experiences can be intricate, as individuals vary in their capacity to recognize, interpret, and articulate their feelings~\cite{salovey1990, lane1990levels, luminet2004multimodal}. For some, especially in psychotherapy, crafting a narrative that describes one's life journey can be a challenging process~\cite{donnelly1991cognitive, pennebaker1999forming}. 

\red{In this work, we aim to lower the barrier of journaling for patients by allowing them to carry on casual conversations with an AI instead of plain open-ended text writing. We demonstrate the potential of these dialogues as a source of PGHD in clinical settings to facilitate individual self-reflection and enhance communication between patients and MHPs.}

\subsection{Conversational Agents for Mental Health}
The field of AI has proposed significant innovations in medical settings, such as aiding clinical decision-making and diagnosis~\cite{magrabi2019artificial, beam2016translating}. \red{In a mental health domain, Natural Language Processing (NLP) techniques have been widely applied to venues that demand human language comprehension and generation for patient support and treatment~\cite{calvo2017natural, demner2009can}.} Since caring mental health often involves counseling, conversational agents, also known as chatbots, have particularly stood out in the mental health domain~\cite{abd2021perceptions, ABDALRAZAQ2019103978}. Studies have demonstrated the potential of chatbots in facilitating different types of therapy, such as cognitive behavioral therapy~\cite{woebot, 10.2196/12106, fulmer2018using}, expressive writing~\cite{park_chatbot}, behavioral reinforcement~\cite{10.2196/12106}, and solution-focused therapy~\cite{fulmer2018using}. Prior work has also shown that chatbots could ease the burden of disclosing sensitive information. Studies indicated that individuals may feel more comfortable communicating with chatbots because of the social stigma involved in communicating with human beings~\cite{10.1145/3313831.3376175, park_chatbot, 10.1145/3392836}. Furthermore, these approaches can help overcome temporal and spatial constraints, offering mental health support that is accessible at any time and anywhere~\cite{cameron2019assessing}. 

Early mental health therapy chatbots predominantly relied upon rule-based or retrieval-based approaches~\cite{ABDALRAZAQ2019103978, abd2019overview}. While these approaches provide a high level of control over the conversational flow, they fall short of carrying on open-ended conversations; that is, the chatbots neither respond to serendipitous topics nor offer versatile responses that are outside the scope of design~\cite{Huang2020ChallengesODD, Gao2018NeuralODD, lee_caring_2019}. In the mental health context, these limitations may impact the quality of the chatbot's caring behaviors as their messages tend to be general~\cite{park_chatbot, Huang2020ChallengesODD}. Recent LLMs have suggested a new paradigm of bootstrapping chatbots ~\cite{chatbot_review21, wei2023leveraging}. LLM-driven chatbots tend to produce human-like, context-aware responses for unseen topics~\cite{wei2023leveraging, ChatGPT}. Consequently, LLM-driven chatbots excel in facilitating open-domain dialogues and engaging in unscripted interactions~\cite{bae2022building}, offering flexibility and adaptability in conversations~\cite{radford2019language, 10.3758/s13428-020-01531-z}, especially in complex scenarios like mental health support. For instance, GPT-3.5 demonstrates empathetic traits, such as recognizing emotions and offering emotionally supportive replies in various situations, predominantly in healthcare settings~\cite{sorin2023large}. \changed{In certain cases, these models appear to have potential in tasks that involve empathy, showing promising results when compared to humans~\cite{ayers2023comparing, elyoseph2023chatgpt}. Given these early findings, LLM-driven chatbots in the public health sector may offer some support in alleviating emotional burden and loneliness among isolated individuals, though this area is still under exploration~\cite{Jo2023CareCallInterviews}. Additionally, the integration of LLMs with human efforts in creating mental health peer support messages could potentially lead to more empathetic conversations~\cite{sharma2023human}.}

However, LLMs suffer from inherent challenges tied to their transformer-based architecture~\cite{vaswani2017attention}. One key issue is the explainability of the model output: it is challenging to discern how this `black box' model interprets a given input prompt. As a result, designers struggle to predict both the LLM's understanding of the input and the subsequent messages it might produce~\cite{liu2023pre}. For instance, a chatbot leveraging GPT-2 for mental health therapy occasionally generated non-word sentences or produced more negative outputs than positive ones~\cite{chatbot_review21}. Replika, an LLM-driven application intended for mental well-being, has occasionally displayed harmful content and exhibited inconsistent conversational styles, undermining its role as a long-term companion~\cite{10.48550/arxiv.2307.15810}.

\red{\changed{These findings highlight two implications for LLM-infused systems in clinical settings. First, human oversight and moderation are critical when using LLMs in clinical settings~\cite{chatbot_review21}. To ensure safe and ethical instrumentation of LLMs for mental health patients, this work involves MHPs to reflect domain experts' perspectives~\cite{Torous2023DigitalResearchPriorities} in developing an LLM-driven chatbot system. Second, it is necessary to enhance the controllability of an LLM~\cite{Jo2023CareCallInterviews, Pereira2023HerdingAICats} to better follow the intended conversational design. As one exemplar approach to enhance LLM's controllability, yet not a conversational domain, AI Chains break down a complex task into multiple, simpler sub-tasks so that individual LLM inferences can work in better reliability~\cite{wu2022aichains}. Similarly, in our system, the chatbot operates using state-based prompting, where a model prompt contains the instruction focused solely on the current state, which is part of an overarching conversational protocol. By simplifying the model prompts, we intended the generated responses to safely and reliably comply with the MHP's guidelines for interacting with the patients.} 

In summary, our work leverages LLMs in two key components: First, we use an LLM to power a chatbot for patient journaling. Second, the clinician dashboard incorporates LLMs for various NLP tasks, such as text summarization and classification, to visualize summarized insights~(\eg,~\cite{Arakawa2023CatAlyst, Gebreegziabher2023PaTAT}) that are noteworthy for MHP's treatment.} \changed{Through this collaborative approach involving clinicians, our work explores a unique design space for designing deploying LLM-driven chatbots in the mental health domain, aiming to improve communication between patients and providers by facilitating the recording of daily experiences, which act as a bridge for a better understanding of patients.}
    \section{Formative Study: Focus Group Interview}
To inform the design of MindfulDiary, we first conducted a Focus Group Interview (FGI) with MHPs. The goal of the FGI was to understand MHPs' perspectives, expectations, values, and precautions in utilizing LLMs in the clinical mental health context. Based on this understanding, we aimed to design the functions and interactions that the system should provide. This was an essential process in our overall approach, not just technology-centered system design, but creating a system meaningful to users and stakeholders~\cite{Thieme2022AIMentalHealth}.

\subsection{Procedure and Analysis}
\changed{We distributed recruitment flyers in the Department of Psychiatry at a local university hospital, inviting Mental Health Professionals (MHPs) working in departments of psychiatry and mental health care centers to participate.} We recruited six MHPs (E1--6; two males and four females)---four clinical psychologists and two psychiatrists whose careers varied from 1 to 11 years. Four were clinical psychologists responsible for counseling and daily monitoring and intervention of at-risk patient groups in local mental health centers and university hospitals, and two were psychiatrists in charge of outpatient and inpatient ward treatment in the psychiatry department of university hospitals (see \autoref{tab:FGI}).

We invited participants to two 1-hour remote sessions on Zoom. Two researchers participated in the sessions. We first provided an overview of language model technologies and LLM's natural language understanding and generation capabilities until we shared a common understanding of the principles, applications, opportunities, and limitations of LLMs. Considering that we were designing a system for individuals with mental health challenges, we thoroughly covered the drawbacks of LLMs, such as uncertainty in control and hallucinations.

\red{After the overview, we went through group discussions on how LLMs could be utilized in the current patient treatment process. As a probe, we asked participants a focused set of questions on (1) the challenges MHPs currently face during patient treatment and counseling sessions, and (2) their expectations and envisioned opportunities of LLMs' role in clinical mental health settings. We sought to understand the experts' perspectives through questions such as, `\textit{What are the difficulties or challenges patients face in their daily lives between treatments (or counseling)?}', `\textit{What are the important considerations in self-care that patients perform in their daily lives?}', and `\textit{What questions or conversational techniques do you use to encourage patients to share about their daily lives and moods?}'. The session was video recorded and later transcribed. We open-coded the transcripts to identify emerging themes. In the following, we cover the findings from the FGI.}

\begin{table}[t]
\sffamily\small
\def\arraystretch{1.2}\setlength{\tabcolsep}{0.4em}
		    \centering
\caption{\red{Demographic of FGI Participants (E1--6).}}
\Description{The table titled "Demographic of FGI Participants (E1--6)" presents data about six participants, coded E1 to E6, in a focus group interview. The table is structured with five columns and seven rows, including the header row. The header row, shaded in light gray, contains the column titles: "Code," "Gender," "Age," "Job Title," and "Years of Experience.}
\label{tab:FGI}
\begin{tabular}{|l!{\color{gray}\vrule}l!{\color{lightgray}\vrule}l!{\color{lightgray}\vrule}l!{\color{lightgray}\vrule}l|}
\hline
\rowcolor[HTML]{EFEFEF} 
\textbf{Code} & \textbf{Gender} & \textbf{Age} & \textbf{Job title}    & \textbf{Years of Experience} \\ \hline
\textbf{E1}            & F               & 34           & Clinical psychologist & 2 years                           \\ \arrayrulecolor{tablegrayline} \hline
\textbf{E2}            & F               & 30           & Clinical psychologist & 1 year                           \\ \hline
\textbf{E3}            & F               & 38           & Clinical psychologist & 9 years                           \\ \hline
\textbf{E4}            & F               & 36           & Clinical psychologist & 11 years                          \\ \hline
\textbf{E5}            & M               & 35           & Psychiatrist          & 2 years                           \\ \hline
\textbf{E6}            & M               & 40           & Psychiatrist          & 10 years                          \\ \arrayrulecolor{black}\hline
\end{tabular}
\end{table}

\subsection{Findings from the Interviews}
\subsubsection{Challenges in Eliciting Responses from Patients with Depression}
Participants indicated that eliciting disclosure from patients' inner thoughts during a limited consultation time requires significant effort. Many patients with depression experience difficulty describing and expressing their feelings and thoughts to providers due to a sense of apathy, which is a common psychiatric symptom involved in Major Depressive Disorder: \textit{``In the consultation room, even if they sit like this, they often just remain silent for a long time.''} (E5)  Thus, providers often end up spending a substantial amount of time asking standardized and repetitive questions about mood, sleep, and major events to understand patients' current states. 

Participants also noted that they had their patients engaged in paper-based diary writing methods but most demonstrated low participation rates and low engagement: \textit{``We tried a diary method on paper(in the inpatient ward), and several patients did write. What we saw was quite trivial, like, `I just felt bad today.' But we learned there were significant events upon consultation, like having a big argument with other patients, which they did not record. Because patients with depression, or those who have had suicidal or self-harming incidents, often have a dulled state in expressing their emotions or feel apathetic, they tend to find such expressions very difficult.''} (E3)

\subsubsection{LLMs as a Bridge for Enhanced Patient Communication}
Our participants envisioned LLMs as a transformative tool in mental health care, particularly for enhancing interactions with patients who struggle to express themselves. They recognized that the natural and flexible conversational abilities of LLMs could bridge communication gaps, offering a more nuanced understanding of patients' conditions. This could be particularly beneficial in cases where patients have difficulty articulating their feelings due to symptoms like apathy or social phobia. Additionally, participants noted that using LLMs could be significantly more interactive and engaging than traditional paper-based approaches, potentially increasing compliance and participation in the therapeutic process. 

\changed{Participants especially underscored the importance of capturing the continuum of thoughts leading up to a particular emotional state, such as fear, in the journaling process. They envisioned the need for using LLMs to introspect deeper into the patient's psyche, revealing underlying thoughts and emotions that the patient might not be consciously aware of. E5 mentioned, "It would be good if the journal continuously records the flow of thoughts. For example, it would be beneficial to document the various thoughts and detailed reasons leading up to certain feelings like fear. Like, 'I feel scared when I'm in a place with many people,' and then digging deeper into 'Why do I feel scared?'---I think a process that gets more specific like this would be good." This approach not only aids in a more comprehensive self-examination but also enriches the therapeutic dialogue between the patient and the MHP.}

\subsubsection{LLMs for Analytical Insights and Personalized Mental Health Support}
\changed{The participants further suggested that LLMs could analyze journal entries to identify key themes, words, or sentiments expressed over time, offering patients tangible feedback on their emotional patterns and progress. Such analytical capabilities could empower patients with a greater sense of control and awareness of their mental health journey, potentially motivating them towards self-management and active participation in their treatment. Additionally, the analysis could assist MHPs in a deeper understanding of their patient's emotional states and thought processes by examining the tone, choice of words, and speech or writing patterns. The participants envisioned that insights derived from LLMs about patient journaling habits could inform MHPs about the most effective counseling approaches for each individual. They suggested, ~\textit{"Observing how patients react to different forms of communication can provide valuable information. Some patients might find solace in simple reassurance, while others may benefit from more straightforward, targeted feedback."}}

\subsection{Improvements after the Interviews}
Based on the lessons from the FGI, we refined the initial concept of MindfulDiary. We leveraged the conversational abilities of LLMs to help patients document their daily experiences between clinical visits. MHPs had access to the collected data to inform their clinical decision-making. Furthermore, both MHPs and the research team concur that LLMs should not act solely as the primary intervention due to their inherent limitations but should function as supportive tools for clinical consultations. The subsequent section outlines the design and development process of our system.
    \begin{figure*}[htbp]
\begin{center}
    \includegraphics[width=\textwidth]{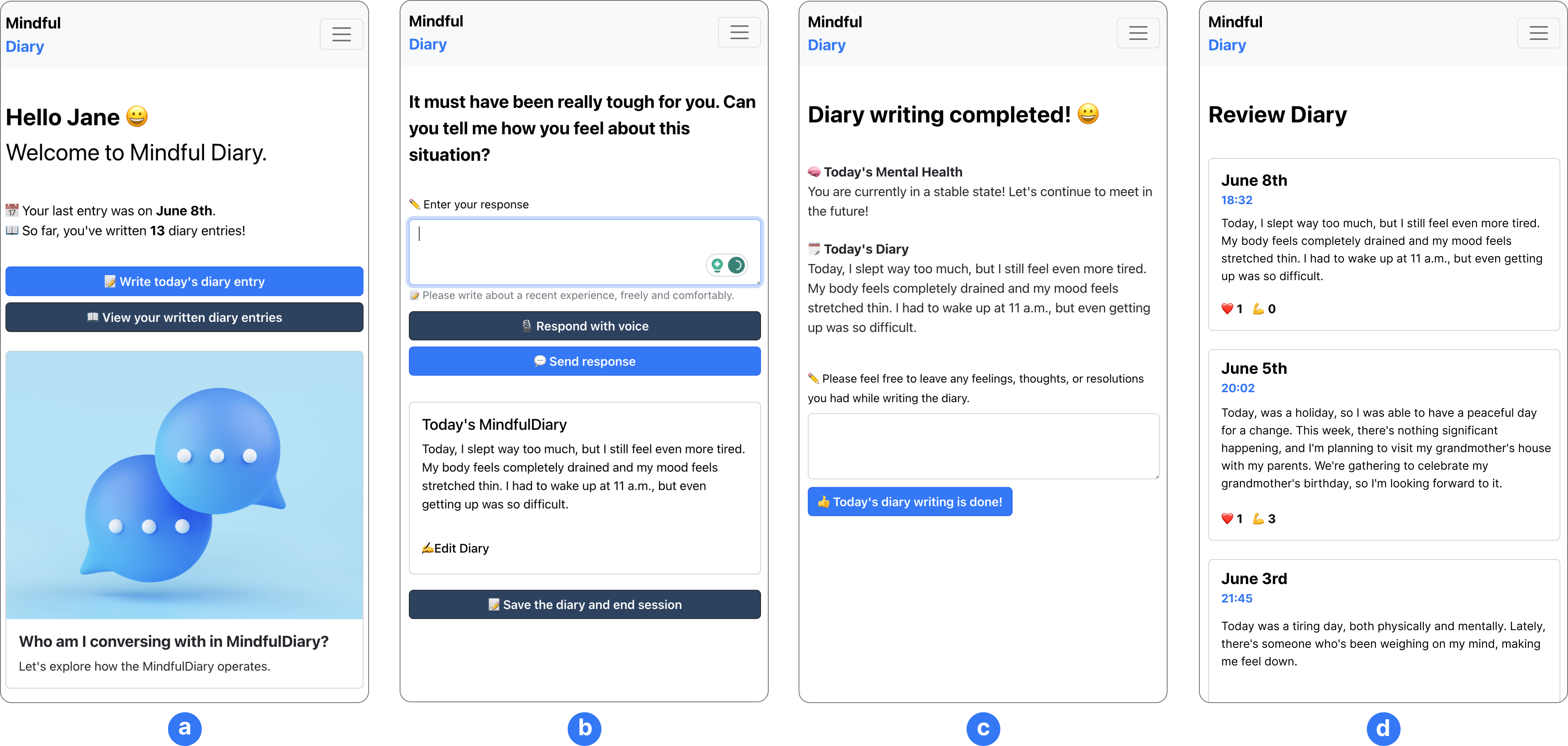}
    \caption{Main screens of the MindfulDiary app. (a) The main screen, (b) the journaling screen, (c) the summary screen shown when the user submitted the journal dialogue, and (d) the review screen displaying the user's past journal.}
    \Description{Composite screenshot displaying four representative screens of the MindfulDiary patient interface. Starting from the far left: (a) A home screen featuring the user's name, recent journal entries count, and a button to initiate a session. (b) A screen captured during a journal session, showcasing a prompt question provided by MindfulDiary, a space for user's response, and a summarized journal entry. (c) A post-journaling summary screen displaying the user's current mental health status, the summarized journal entry, and a space for personal reflection. (d) A review screen where users can revisit their past journal entries, with entries visualized as individual cards.}
    \label{fig:design:screens}
    \labelphantom{fig:design:screens:main}
    \labelphantom{fig:design:screens:journaling}
    \labelphantom{fig:design:screens:summary}
    \labelphantom{fig:design:screens:review}
\end{center}
\vspace{-4mm}
\end{figure*}

\section{MindfulDiary} 
Informed by the findings from FGI with MHPs, we designed and developed MindfulDiary, which consists of two main components: (1) a patient mobile app for daily record-keeping and (2) a clinician dashboard that allows professionals to access and use these daily records in a clinical setting~(See \autoref{fig:teaser}). Below, we present a fictional usage scenario to demonstrate how the system works.


\paragraph{
\textit{Jane, diagnosed with chronic anxiety, frequently grapples with panic attacks. To keep track of her daily experiences, her psychiatrist recommends trying MindfulDiary as part of her treatment plan.}}

\paragraph{\textit{Every evening, Jane converses with the MindfulDiary app regarding her daily activities, emotions, and thoughts. \changed{The AI leads the conversation with Jane by asking prompted and follow-up questions about her day.} After a session, the app summarizes the dialogue into a journal-style essay, on which she can revisit and reflect later. She can explore the summarized essays whenever she wants to reflect on past events or thoughts.}}
        
\paragraph{\textit{Three weeks later, during a consultation, her psychiatrist uses the expert interface of MindfulDiary to review a data-driven summary of Jane's entries. The data helped the psychiatrist identify patterns that Jane's anxiety often spikes during her work commute. Based on this insight, the psychiatrist refines advice and introduces specific coping strategies, fostering a more personalized approach to care.}}

\subsection{MindfulDiary App}
The MindfulDiary app for patients aims to support people who might have difficulty journaling due to apathy and cognitive load through naturalistic conversation driven by an LLM. The app consists of a home screen containing an introduction and guide to the system~(\autoref{fig:design:screens:main}), a journal writing screen (\autoref{fig:design:screens:journaling}, \ref{fig:design:screens:summary}), and a screen to review the diary entries~(\autoref{fig:design:screens:review}).

\subsubsection{Journaling User Interface}

\begin{figure*}[t]
\begin{center}
    \includegraphics[width=\textwidth]{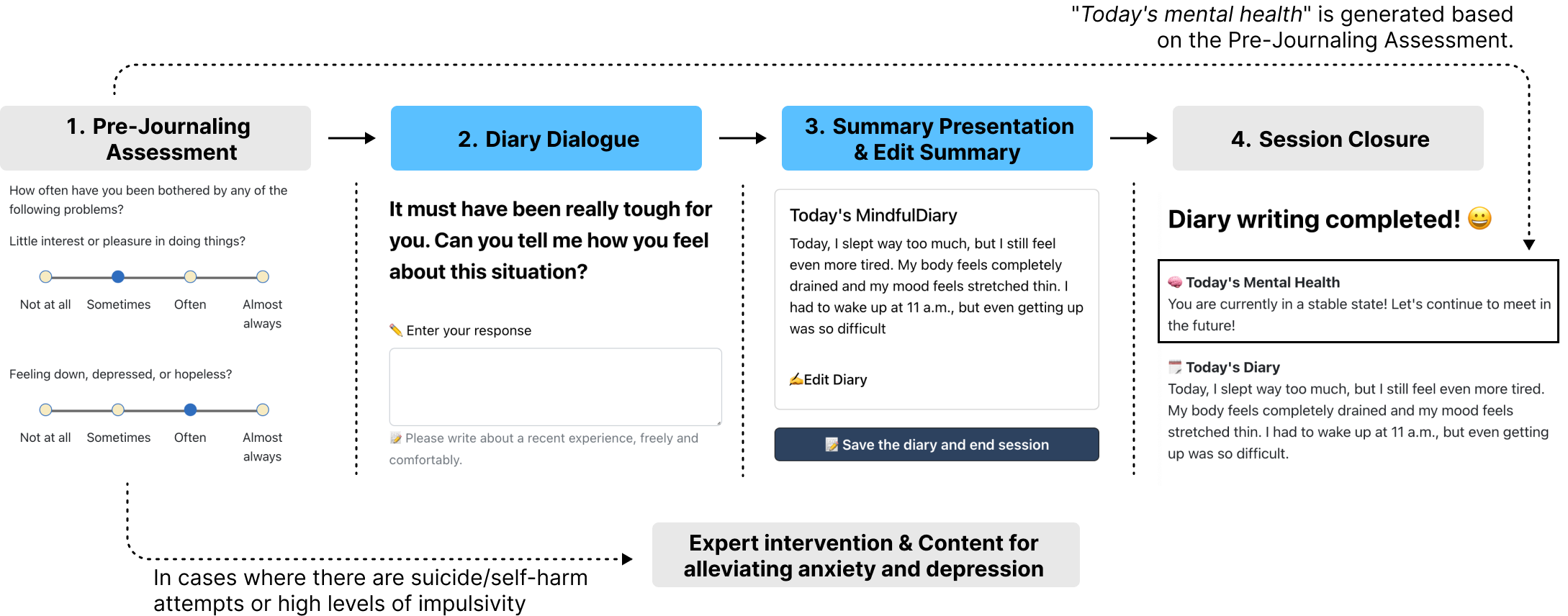}
    \caption{Use flow of MindfulDiary's journaling session: (1) Pre-Journaling Assessment: Users undergo a mental health survey using the modified PHQ-9~\cite{kroenke2001phq} before using MindfulDiary; (2) Users converse with  MindfulDiary, documenting their day; (3) Summary Presentation: After three turns, MindfulDiary presents a diary-styled summary of the conversation so far, which can also be edited by the user. Users can continue the conversation as they want. (4) Session Closure: Once all processes are completed, MindfulDiary displays today's mental health and diary content, concluding the journaling session.}
    \Description{Sequential visual representation of the MindfulDiary's journaling session flow, depicted from left to right across four cards: (1) 'Pre-Journaling Assessment': Illustration of users taking a modified PHQ-9 mental health survey prior to their session with MindfulDiary. (2) 'User-MindfulDiary Conversation': Image of users engaging in a conversation with MindfulDiary, narrating and documenting their day. (3) 'Summary Presentation': Graphic depicting a diary-style summary of the three-turn conversation, with an option for users to edit. It signifies users' choice to either continue or end the conversation. (4) 'Session Closure': Visualization of the session's end screen, where MindfulDiary showcases the day’s mental health assessment alongside the diary content, signifying the conclusion of the journaling experience.}
    \label{fig:useflow}
\end{center}
\end{figure*}

\autoref{fig:useflow} illustrates the overall use flow of the journaling session, which begins with a Pre-Journaling Assessment (\autoref{fig:useflow}-\circledigit{1}) that asks to fill out a questionnaire for mental health. The questionnaire comprised the modified PHQ-9~\cite{kroenke2001phq} and a custom open-ended question inquiring about recent attempts of self-harm or suicide.
This assessment prevents users who provided any clues of suicidal or self-harm from journaling on the same day. (We cover this feature in detail in \autoref{sec:study:ethical}.)

On the next screen, the user converses with MindfulDiary, documenting the events of the day~(See \autoref{fig:design:screens:journaling}). After three turns, MindfulDiary provides a summary of the conversation as an essay. Users can edit this automatically generated summary any time. When the user ends the session by pressing the end button (\autoref{fig:design:screens:journaling}, bottom), MindfulDiary displays daily mental health insights alongside the diary content on the summary screen (See \autoref{fig:design:screens:summary}). Users can also leave a reflection message there. Lastly, users can browse their past records in the Diary Review menu (See \autoref{fig:design:screens:review}).

\subsubsection{Conversation Design}
We designed the chatbot's conversational behavior based on insights from psychiatry literature~\cite{othmer2002, clinical1}, which covers foundational techniques and considerations for conducting clinical interviews. We also incorporated the hands-on clinical experiences of practicing psychiatrists.

As a result, we designed the conversation of a journaling session to follow a sequence of three stages: \textit{Rapport building}, \textit{Exploration}, and \textit{Wrap-up}. The \textbf{Rapport Building} state is an ice-breaker, centered on casual exchanges about a user's day. In this state, the assistant also shares bits of information to encourage users' openness. \red{This approach is based on previous research findings that a chatbot's self-disclosure positively impacts user disclosure~\cite{10.1145/3392836} and leverages the natural story-building ability of LLMs~\cite{10.1145/3591196.3596612}. Overall, in this stage, our goal is to create an environment where users can comfortably share their stories.} As we progress to the \textbf{Exploration} state, the emphasis shifts to a comprehensive understanding of the user's daily events, feelings, and thoughts, facilitated by a mix of open-ended and closed-ended queries that ensure users remain engaged and in control of the dialogue.~\red{While open-ended queries are intended to facilitate increase the expression of feelings and emotion and less judgemental, closed-ended queries is for specific and detailed description of the experiences~\cite{othmer2002, clinical1}.}The conversation then transitions to the \textbf{Wrap-up}, emphasizing completion and ensuring users have fully voiced their experiences while the system remains empathetic and receptive to any lingering topics.

Besides the three main stages, we also incorporated the \textbf{Sensitive Topic} state that handles the most sensitive subjects, such as self-harm or suicidal ideation. When this state is triggered, psychiatrists receive instant notifications. This allows them to oversee the conversation in real-time and step in to assist the patient if necessary. Here, the system begins by empathizing with the user, recognizing their struggles, and offering a reassuring message. Following this, the system gently probes the depth of their suicidal or self-harm thoughts. If the user expresses intense or specific plans related to self-harm or suicide, the system urges them to seek prompt assistance, either at a hospital or via the local helpline.

\subsubsection{Conversational Pipeline}

Lengthy and complex input prompts for LLMs are known to cause poor task performance~\cite{brown2020language} by partly omitting latent concepts~\cite{wu2022aichains}. To steer the LLM to comply with the conversational design we intended diligently, we designed MindfulDiary's dialogue system as a state machine. Each conversation stage is carried on with a dedicated input prompt, which is more succinct and clear than a single master prompt containing instructions for all stages.

\begin{figure*}[t]
\begin{center}
    \includegraphics[width=\textwidth]{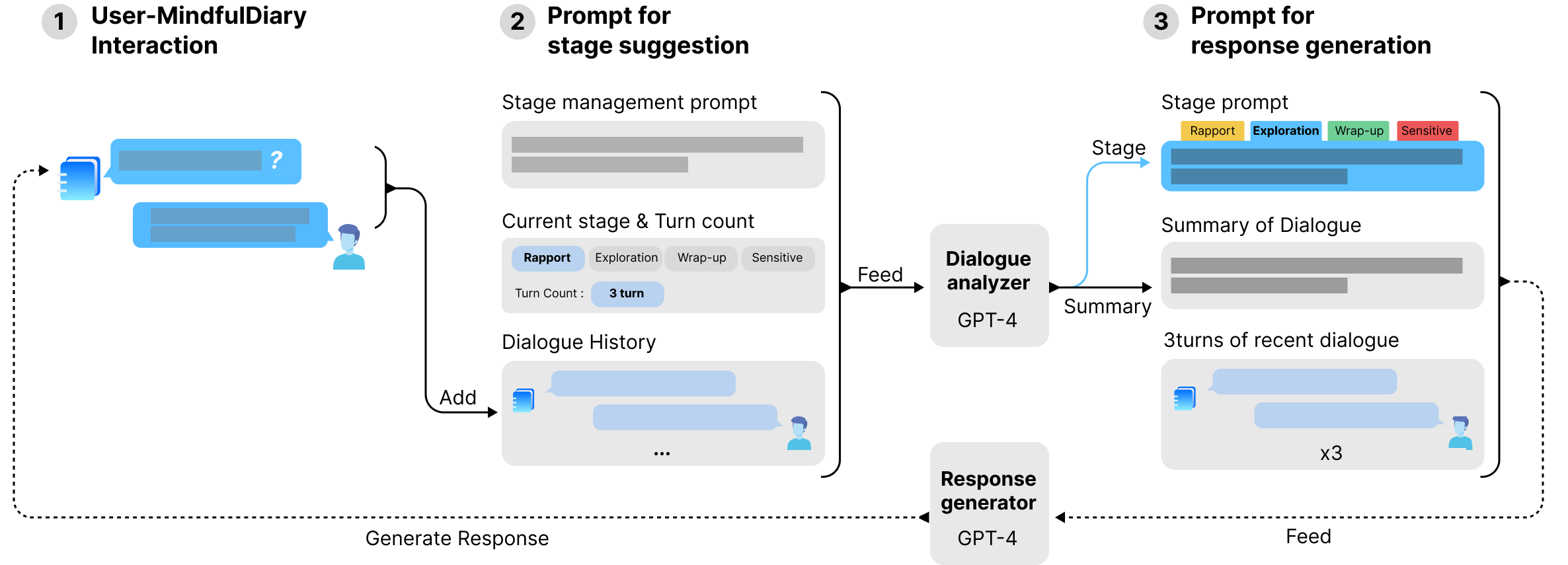}
    \caption{Structure of MindfulDiary's conversational pipeline. (1) Users respond to MindfulDiary's messages. (2) The recent turn count, current state, and whole user-MindfulDiary dialogue are fed into Dialogue Analyzer. (3) Using the output of Dialogue Analyzer, a designated state prompt, a summary of dialogue (containing overall dialogue context), and the latest three conversation turns are fed into Response Generator. The resulting response is then displayed to the user. Both the Dialogue Analyzer and Response Generator operate based on the GPT-4 LLM.}
    \Description{Visual representation of the MindfulDiary's conversational prompt pipeline, segmented into three distinct modules from left to right: (1) 'User Interaction': Depicts users engaging with and responding to MindfulDiary's messages. (2) 'Dialogue Analyzer': Shows the intake of the recent turn count, current state, and the entirety of the user-MindfulDiary dialogue. This module processes the conversation dynamics and is built upon the GPT-4. (3) 'Response Generator': Illustrates how the results from the Dialogue Analyzer, alongside a specified state prompt, summarization of the overall dialogue context, and the most recent three turns of the conversation, are inputted to produce a fitting response. This generated response is subsequently presented to the user. This module also operates on the GPT-4.}
    \label{fig:prompt}
\end{center}
\end{figure*}

\autoref{fig:prompt} illustrates our conversation pipeline that runs each time a new user message is received. The pipeline incorporates two LLM-driven modules, a \textit{dialogue analyzer} and the \textit{response generator}. 

The \textbf{dialogue analyzer} handles the stage transition, returning the stage suggestion---whether to stay or move to a new stage---and a summary paragraph of the current dialogue from the current dialogue information. \red{The dialogue analyzer receives an input prompt that consists of the current number of turns in the conversation (progress level), the most recent stage information, and a list of criteria for recommending each stage (See \circledigit{2} in \autoref{fig:prompt}). Based on this information, the underlying LLM performs a summarization task that yields a summarized paragraph of the current dialogue, a recommendation for the next stage based on the summary, turns, and the most recent stage information. For example, the system decides to move to the Wrap-up stage when the user expresses a desire to conclude or say goodbye.}

\changed{The system then formulates an LLM prompt, combining a dedicated prompt for the current stage, the dialogue summary, and the recent six messages (\ie, three turn pairs)~(See \circledigit{3} in \autoref{fig:prompt}).} Receiving the prompt as an input, the \textbf{response generator} generates an AI message. \red{The stage prompt consists of the description of the task that the LLM is supposed to perform in the current stage, and the speaking rules describing the attitude that the module exhibits in the conversation. For example, the task description of Exploration stage instructs to ``\texttt{ask questions that encourage users to reflect on their personal stories regarding daily events, thoughts, emotions, challenges, and etc.}''  The speaking rules for the Rapport-building stage instruct to keep conversations simple and friendly and reply in an empathetic way.}

\subsection{Pilot Evaluation}
To ensure that MindfulDiary is reliable and safe for conversing with psychiatric patients, we underwent multiple rounds of pilot evaluation. First, we invited five psychiatrists and three clinical psychologists to test the conversational pipeline. The experts provided feedback on the instructions in the model prompts, focusing on their clinical relevance and the embedded terminology and strategies. Then, the experts inspected the chatbot's behavior by chatting with it while role-playing as a patient persona. In particular, we examined the chatbot's reactions to subtle implications of suicide or self-harm in user messages. 

After iterating on the conversational pipeline, we conducted a pilot lab study with five patients admitted to a university hospital but about to be discharged soon. To ensure safety against risky messages from an LLM, we used a test platform where the participant's clinician monitored the generated messages in real-time, approving them or sending better messages manually.

\subsection{Clinician Dashboard}
The clinician dashboard (\cf, Supplementary video) is a desktop application designed to facilitate monitoring patient's journal entries and to provide analysis of the entries to help clinicians identify significant events, reactions, and emotions. The dashboard consists of the following components:

\ipstart{User Engagement} This section visualizes the participant's overall engagement with MindfulDiary, including the number of journals written, the date and time they were written, and their length. The modified PHQ-9 scores for each session are also visualized, allowing professionals to track the user's mental health trends using a validated tool.

\ipstart{Journals} This section displays the content of the journals written by patients. The information is presented in a card format, where each card offers a summary of the journal, including timestamps, total time taken to write the journal, and associated PHQ-9 score. The  interaction logs between the patient and MindfulDiary are also provided in this section.

\ipstart{Insights} To assist professionals in browsing through the diary, this section visualizes (1) a word cloud to understand frequent terms that the participant used at a glance, (2) a summary of major events to highlight significant happenings and (3) summary of emotions to gauge the mood based on user input. When a specific period is selected for review, a comprehensive summary is generated. \red{We used GPT-4 for most summarization tasks. To generate the word frequency data for the word cloud, we combined GPT-4 and a Korean morphological analysis package named Kiwi~\cite{minchul_lee_2022_7041425} to filter only nouns and verbs from the GPT output.} Due to the limitations of language model-driven analysis, there might be occasional inaccuracies in the generated content. First-time users of this interface are alerted about possible inaccuracies. An in-interface tooltip also reminds users that the summarized outcomes might not be accurate.

\subsection{Technical Implementation}
MindfulDiary's interface is developed using React, a JavaScript-based framework. The server, responsible for interfacing with the LLM and overseeing database operations, is implemented in Python. Google Firebase handles user authentication, data storage, and retrieval tasks. The conversational capabilities of MindfulDiary are powered by \textit{gpt-4}, accessible through OpenAI's API\footnote{https://platform.openai.com/docs/guides/gpt/chat-completions-api}. We specifically used \texttt{gpt-4-0613} model. For parameter setting, we consistently set the temperature to 0.7 and both a presence penalty and frequency penalty to 0.5.

    \section{Field Deployment Study} 

Using MindfulDiary, we conducted a four-week field deployment study with 28 patients undergoing outpatient treatment. Through the study, we aimed to explore how patients and MHPs utilize MindfulDiary and what opportunities and challenges arise from its real-world use. \red{The study protocol was approved by the Institutional Review Board of a university hospital.} 

\subsection{Recruitment}

We targeted outpatients from the Department of Mental Health at a University Hospital. Participants were selected based on specific criteria: (1) those who had been diagnosed with MDD and (2) those who did not exhibit heightened impulsive tendencies or harbor specific intentions towards self-harm or suicide. \red{Key exclusion criteria included a history of psychotic disorders, substance-related disorders, neurodevelopmental disorders, and neurological disorders.} Eligible participants were identified through evaluations conducted by psychiatrists. Flyers and consent forms were distributed to eligible patients. For minors, the consent form process was adhered to only when they were accompanied by a guardian at the hospital.

\begin{table}[b]
\sffamily\small\def\arraystretch{1.2}\setlength{\tabcolsep}{0.4em}
		    \centering
\caption{\red{Demographic of MindfulDiary participants (P1--28). The table presents gender, age, and the severity of depressive symptoms represented by CES-DC (Center for Epidemiologic Studies Depression Scale for Children), which assesses symptoms of depression in children and adolescents. Scores were categorized as follows: below 16 as \cesminimalbox{Minimal}, 16 and above as \cesmildbox{Mild}, and 25 or higher as \cesseverebox{Severe}~\cite{cho2001prevalence}.}}
\Description{The table titled "Demographic and Psychological Assessment Scores of MindfulDiary Participants" presents the gender, age, and severity of depressive symptoms using the CES-DC scale for 28 participants, labeled P1 to P28. The table has four columns: "Code," "Gender," "Age," and "Severity of depressive symptoms (CES-DC)." The CES-DC scale categorizes scores as "Minimal" (below 16), "Mild" (16 and above), and "Severe" (25 or higher).}
\label{tab:demographics}
\begin{tabular}{|c!{\color{gray}\vrule}c!{\color{tablegrayline}\vrule}c!{\color{tablegrayline}\vrule}c|}
\hline
\rowcolor{tableheader} 
\textbf{Alias} & \textbf{Gender} & \textbf{Age} & \textbf{\begin{tabular}[c]{@{}c@{}}Severity of depressive symptoms (CES-DC)\end{tabular}} \\ \hline
\textbf{P1}           & F              & 16           & \cesminimalbox{Minimal}                                           \\
\arrayrulecolor{tablegrayline}\hline
\textbf{P2}           & F              & 19           & \cesmildbox{Mild}                                              \\
\hline
\textbf{P3}           & M              & 17           & \cesminimalbox{Minimal}                                           \\
\hline
\textbf{P4}           & F              & 17           & \cesseverebox{Severe}                                            \\
\hline
\textbf{P5}           & F              & 14           & \cesseverebox{Severe}                                            \\
\hline
\textbf{P6}           & F              & 17           & \cesseverebox{Severe}                                            \\
\hline
\textbf{P7}           & F              & 16           & \cesminimalbox{Minimal}                                           \\
\hline
\textbf{P8}           & F               & 14           & \cesmildbox{Mild}                                              \\
\hline
\textbf{P9}           & M               & 16           & \cesmildbox{Mild}                                              \\
\hline
\textbf{P10}           & F               & 16           & \cesseverebox{Severe}                                            \\
\hline
\textbf{P11}           & F               & 19           & \cesseverebox{Severe}                                            \\
\hline
\textbf{P12}           & M               & 19           & \cesseverebox{Severe}                                            \\
\hline
\textbf{P13}           & F              & 19           & \cesmildbox{Mild}                                              \\
\hline
\textbf{P14}           & M               & 18           & \cesseverebox{Severe}                                            \\
\hline
\textbf{P15}           & M               & 24  & \cesminimalbox{Minimal}                                           \\
\hline
\textbf{P16}           & F               & 12           & \cesseverebox{Severe}                                            \\
\hline
\textbf{P17}           & F               & 15           & \cesseverebox{Severe}                                            \\
\hline
\textbf{P18}           & F              & 17           & \cesmildbox{Mild}                                              \\
\hline
\textbf{P19}           & F               & 23  & \cesseverebox{Severe}                                            \\
\hline
\textbf{P20}           & M               & 17           & \cesminimalbox{Minimal}                                           \\
\hline
\textbf{P21}           & M               & 19           & \cesseverebox{Severe}                                            \\
\hline
\textbf{P22}           & M               & 17           & \cesminimalbox{Minimal}                                           \\
\hline
\textbf{P23}           & F               & 28  & \cesseverebox{Severe}                                            \\
\hline
\textbf{P24}           & M               & 17           & \cesminimalbox{Minimal}                                           \\
\hline
\textbf{P25}           & M               & 19           & \cesminimalbox{Minimal}                                           \\
\hline
\textbf{P26}           & F               & 19           & \cesseverebox{Severe}                                            \\
\hline
\textbf{P27}           & M               & 15           & \cesmildbox{Mild}                                              \\
\hline
\textbf{P28}           & F               & 14           & \cesseverebox{Severe}                                            \\ \arrayrulecolor{black}\hline
\end{tabular}
\end{table}

\red{We compensated participants on a weekly basis of participation: For participating every seven days from the starting date, participants received 15,000 KRW (approx. 11 USD). If they completed the entire four-week study process, they received 20,000 KRW as a bonus (\ie, 80,000 KRW---approx. 60 USD---in total). We did not tie the number of dialogue entries to the compensation to ensure natural data entry behavior.}

\red{As a minimum requirement for study completion, we instructed the participants not to miss four consecutive days without conversing with MindfulDiary. If a participant missed three consecutive days, an experimenter sent a reminder. In cases where participants did not respond to these reminders, their participation in the study was discontinued. This procedure was implemented to ensure active monitoring and communication. Considering that our system is designed for \changed{individuals with mental health challenges}, it was crucial to maintain contact with participants and ensure their adherence to the study protocol.}

\red{Initially, 36 patients started using MindfulDiary. During the deployment, eight dropped out as they did not meet the minimum data collection requirement. These participants were disengaged from MindfulDiary due to the lack of time or decreased interest.} As a result, 28 participants (P1--28; 11 males and 17 females) completed the 4-week field study and were included in the analysis. The majority of participants were adolescents and adults, with ages ranging from 12 to 28 years, with an average age of 17.6 ($SD=3.26$). \red{\autoref{tab:demographics} presents the demographic details and severity of depressive symptoms of the study participants. These scores are derived from psychiatric evaluations conducted within one week before the starting dates.} 

\subsection{Procedure}
\begin{figure*}[t]
\begin{center}
    \includegraphics[width=\textwidth]{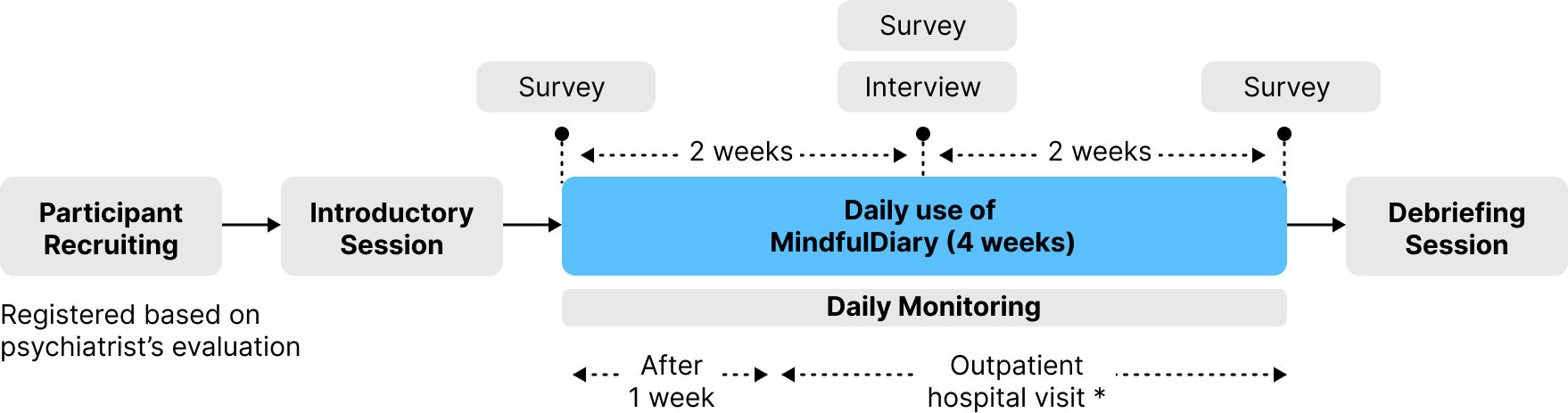}
    \caption{Procedure of the four-week field deployment study: A four-week exploration into the utilization of MindfulDiary by outpatient patients, encompassing daily use, and its integration into clinical decision-making. We note that some participants did not have a follow-up visit during the experimental period. The surveys are outside the scope of this work's investigation.}
    \Description{Timeline diagram illustrating the progression of the field study in the research. Moving from left to right in chronological order: (1) `Participant Recruitment,' (2) `introductory session' (3) `4-week MindfulDiary Daily Use' (4) `Surveys & Interviews: Three distinct markers or symbols at the beginning, middle (after two weeks), and end (after four weeks) of the usage period, highlighting the moments when surveys and interviews were conducted. (5) `researcher monitoring'}
    \label{fig:fieldstudy}
\end{center}
\end{figure*}

\autoref{fig:fieldstudy} illustrates the procedure of the field deployment study. All interviews took place remotely on Zoom.

\subsubsection{MindfulDiary App}
We deployed the MindfulDiary app to our patient participants. The patient protocol consisted of three parts: (1) an introductory session, (2) deployment, and (3) interviews.

\ipstart{Introductory Session}
We first invited each participant to a remote introductory session. A researcher went through our study goal, the motivation of the MindfulDiary system, and the overall procedure of the study. We then played a demo video demonstrating how to use the MindfulDiary app. The session took about 15 minutes.

\ipstart{Deployment}
The day following the introductory session, participants started using MindfulDiary for four weeks. We instructed participants to engage with the app whenever they have anything noteworthy but encouraged them to use it at the end of the day. We collected all data from their interactions with the MindfulDiary and the raw input content and outputs from the LLM. \red{We asked our participants to fill out online surveys three times, at the beginning of Week 1, after Week 2, and after the deployment, to measure participants' mental health status and their self-help capability in managing their mental health. The surveys utilized the PHQ-9~\cite{kroenke2001phq}, GAD-7~\cite{spitzer2006brief}, and Coping Strategies Scale~\cite{zhao2022development}. (The survey results from the scale are outside the scope of this investigation.)}

\ipstart{Mid-study and Debriefing Interviews}
We conducted two 15-minute interviews, after the second and fourth weeks, with each participant to understand their experiences and learn how they used MindfulDiary on a daily basis. Considering the characteristics of depression patients, who may struggle to focus for long periods of time, the interview session was divided into two shorter sessions. 

\subsubsection{Clinician Dashboard}
Most patient participants had a clinical visit during Week 2 through Week 4 of the deployment period. We deployed MindfulDiary's clinician dashboard to five psychiatrists who are in charge of the participants.

\ipstart{Deployment of Clinician Dashboard in Clinic}
We provided instructions to clinicians covering the main components of the clinician dashboard and how to interact with them. To explore the opportunities and limitations of the dashboard, we did not offer explicit instructions for utilizing the clinician dashboard in their workflow. However, we advised psychiatrists to be cautious with the LLM-driven analysis due to potential inaccuracies, emphasizing the importance of verifying data through the interaction logs. The psychiatrists autonomously utilized the clinician dashboard, making sure it didn't disrupt their current treatment methods and preparation routines.

\ipstart{Debriefing Interviews}
We interviewed psychiatrists who treated the patient participants to understand how they used the clinician dashboard in clinical settings. We further gathered feedback from the psychiatrists on the opportunities and limitations of MindfulDiary, as well as suggestions for improvements. The interviews with the psychiatrists were conducted offline for about one hour after the deployment study concluded. 

\subsection{Analysis}
To explore participants' usage patterns with MindfulDiary, we first conducted a descriptive statistics analysis. To determine any shifts in participants' adherence over time, we examined weekly writing frequencies using a one-way repeated measures ANOVA (RM-ANOVA) with Greenhouse-Geisser correction. To gain a deeper qualitative understanding of the messages produced by MindfulDiary and interviews with patients and psychiatrists, we used open coding paired with thematic analysis~\cite{braun2006using}. For a more in-depth qualitative analysis of the messages produced by MindfulDiary and the interviews with patients and psychiatrists, we employed open coding paired with thematic analysis~\cite{braun2006using}. All interviews were audio-recorded and transcribed for this purpose.

\red{The qualitative analysis was conducted by the first author, a PhD student in HCI, who open-coded the interview transcripts and interaction log data through multiple rounds of iteration. Another author who holds a PhD degree in HCI also contributed to this coding process. Following the initial coding, two psychiatrists reviewed the coded data to provide clinical insights and ensure the accuracy of interpretations. Through discussions among the research team, including these diverse perspectives, overarching themes were identified, enhancing the depth and validity of our qualitative findings.}

\subsection{Ethical Considerations}\label{sec:study:ethical}
Conducting this study, we are fully aware of the inherent risks associated with our research, particularly given the characteristics of participants diagnosed with MDD. To mitigate the risks, we first carefully screened participants, relying on evaluations conducted by psychiatrists. Individuals displaying heightened impulsive tendencies or harboring specific intentions towards self-harm or suicide were excluded from the study. In addition, participants were asked to take the PHQ-9 before interacting with MindfulDiary, along with an additional set of questions probing their recent attempts at self-harm or suicide. If a participant's response to question number 9 of the PHQ-9, regarding suicidal/self-harm thoughts, scored `moderate or higher' or if any recent suicide attempt was verified, the system pivoted to provide content geared towards alleviating anxiety and reducing stress rather than proceeding with the standard system. In such a case, a real-time alert was also sent to psychiatrists. Lastly, if sensitive themes frequently surfaced in a participant's input during the study, their interactions with the system were temporarily halted. Psychiatrists subsequently re-evaluated such participants to assess the viability of their ongoing participation. During our experiment, for the case of P11, mentions of repetitive suicide and self-harm were detected. Consequently, an expert contacted the participant, the experiment was suspended for three days, and after a re-evaluation in an outpatient clinic, we resumed the system use with P11.

Further, to mitigate potential risks from the LLMs' outputs, we embraced an iterative design methodology. The system's interactions underwent repeated assessments to ensure it generated safe, non-harmful outputs. In addition, in the first week of each participant's system use, all interactions between participants and MindfulDiary were observed in real time. To facilitate this process, when a participant started the session, the research team received a notification email. This notification included real-time monitoring links and reports of the survey responses that participants answered before each session. After the first week, user interactions and MindfulDiary were reviewed within a 12-hour window. During the review process, if an interaction contained sensitive content (specifically, terms pre-defined as sensitive by psychiatrists), the psychiatrists on our research team assessed the situation and contacted the affected participants if necessary.

Lastly, given that we were handling the patients' personal and sensitive data, ensuring the secure protection and management of data was critical. Therefore, during the study, we utilized the Google Firebase authentication service to manage the user authentication process for participants. We were thus able to ensure that only authorized personnel had access to the data, and any attempts at unauthorized access could be promptly detected and managed. After the field study, all data was separated from personal identifiers to maintain anonymity.

    \begin{figure*}
    \centering
    \includegraphics[width=\textwidth]{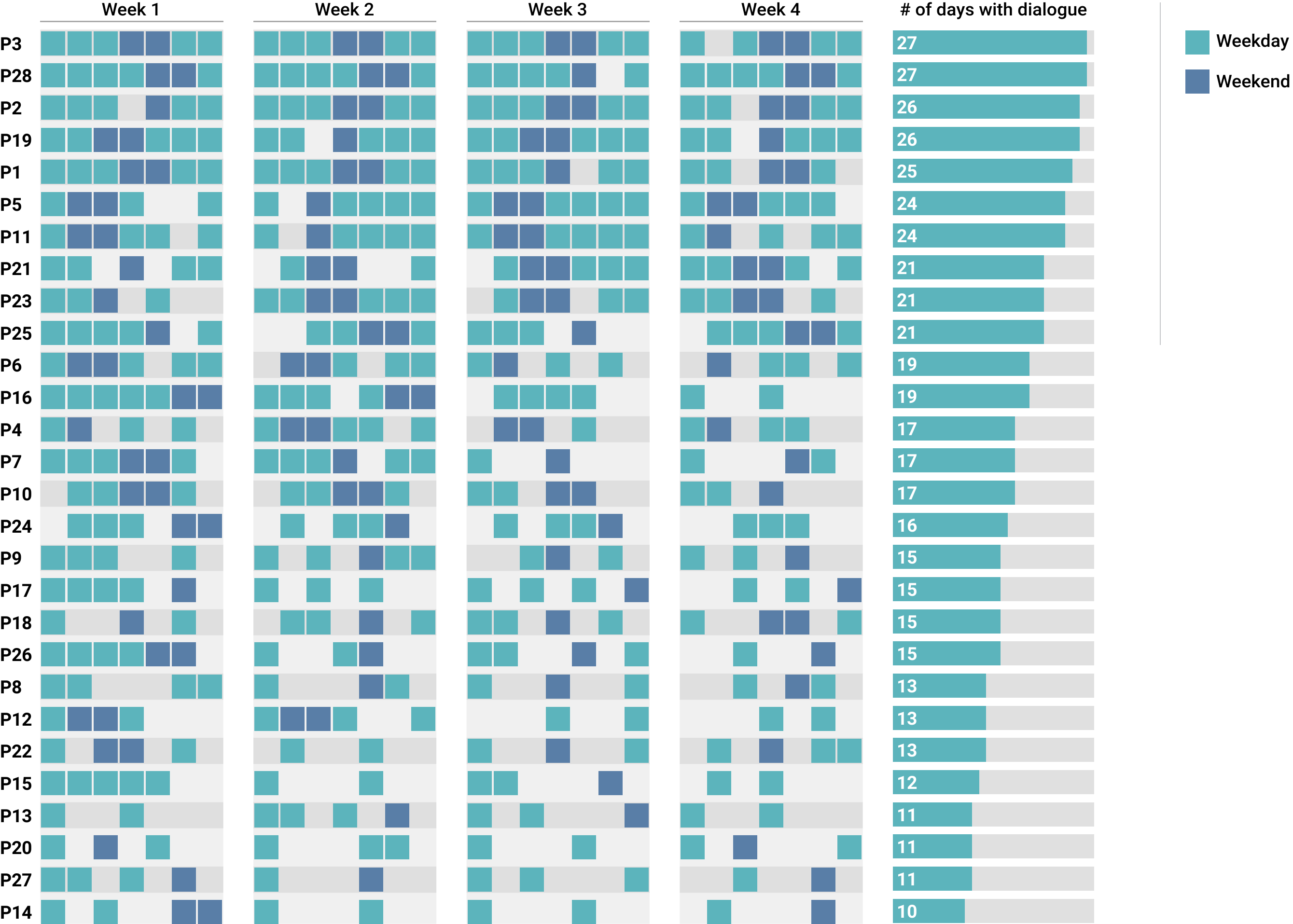}
    \caption{\red{Overview of daily engagement of participants with MindfulDiary. The colored squares denote the days that participants conversed with MindfulDiary, with darker colors indicating weekend days. The bar charts on the right visualize the total number of days with interaction against the four-week study period. Participants are sorted by the number of days with interaction.}}
    \Description{The figure presents an overview of participant engagement with MindfulDiary over a four-week study period. It features a matrix with participants listed on the vertical axis and the days of the study on the horizontal axis. Colored squares indicate days participants interacted with MindfulDiary, with darker shades representing weekends. To the right, bar charts show the total number of days each participant engaged, sorted by their level of interaction. This visualization aims to highlight patterns of usage and identify how engagement varies among participants throughout the study.}
    \label{fig:results:engagement}
\end{figure*}

\section{Results}
In this section, we report the results of the field study in four parts: (1) Journaling adherence, (2) Dialogue patterns, (3) Patients' perspectives on MindfulDiary, and (4) MHPs' perspectives of MindfulDiary for clinical settings.

\subsection{Journaling Adherence}

\red{\autoref{fig:results:engagement} summarizes the daily engagement of participants with MindfulDiary over the course of four weeks. The colored squares denote the days that participants conversed with MindfulDiary (\ie, days with interaction).} Across four weeks, participants submitted 501 journal entries (17.90 entries per participant on average), 0.62 entries on average per day (more than once every two days). \changed{22 out of 28 participants used MindfulDiary more than once every two days.} Participants generally engaged with the app at a regular frequency, but we note that their engagement was also affected by the three-day-miss reminder and their visit to the clinic between Week 2 and 4. Each journaling session lasted an average of 438 seconds (around 7 minutes) but with notable individual variability ($SD=225.97$). Each journal dialogue included messages with an average of 105.6 syllable count ($SD=49.41$). Our analysis did not reveal significant differences in either the participants' input length ($F(1.735, 46.85) = 2.718$, $p=.084$) or writing time ($F(2.417, 65.25) = 2.549$, $p=.076$) across the four different time points, as determined by the RM-ANOVA test. This suggests that users mostly retained a steady level of engagement during the four-week study. 

\subsection{Dialogue Patterns}
Participants and MindfulDiary exchanged a total of 4,410 messages (\ie, 2,205 pairs of the AI and participant's messages) during the field study. Each session consisted of 10.82 messages ($SD=2.70$).
Most exchanges between the AI and participants were carried on for an exploration of patients' daily lives and emotions, as well as for casual conversations. In terms of the stage of the conversation, 62\% (2,732 messages) of the messages were from Exploration, 30\% (1220 messages) for Rapport building, and 6\% (282 messages) for Sensitive topic. Only a small amount of messages were accounted for Wrapping up (62 messages) or not selected (14 messages).

\begin{table*}[!ht]
    \def\arraystretch{1.2}\setlength{\tabcolsep}{0.2em}
    \centering
    \smaller\sffamily
    \caption{Categorization of LLM's strategies for generating prompts to encourage user reflection, including descriptions and exemplar questions. Strategies include Emotional Exploration, Activity Exploration, In-depth follow-up, and Future Plan Exploration with associated descriptions and examples}
    \Description{Table presents a categorization of LLM’s strategies to produce prompts aimed at fostering user reflection. The table is structured into three columns: 'Category', 'Description', and 'Example'. 'Emotional Exploration': This category focuses on probing further into the user’s emotional state, mood, and condition. Sample questions include: “How did you feel after meeting her?” and “Can you tell me how this situation makes you feel?”. 'Activity Exploration': This strategy explores the user’s experiences or undertakings, leaning towards more factual content. For instance, questions such as “What kind of exercise do you do?” and “Lunch with a friend sounds nice. What did you eat?” fall under this category. 'In-depth follow-up': This category dives deeper into the root causes or reasons behind the user's moods, emotions, or particular situations. Example prompts include: “Since when have you felt this overshadowed mood?” and “If you feel confused, what might be the reason?”. 'Future Plan Exploration': Here, the emphasis is on how the user reacted or plans to tackle the events or emotions they've described. Example questions include: “That situation must have been challenging. Have you considered any solutions?” and “What have you done to alleviate the loneliness, even just a bit?}
    \begin{tabular}{|>{\raggedright\let\newline\\\arraybackslash\hspace{0pt}}m{0.15\textwidth}|m{0.40\textwidth}!{\color{tablegrayline}\vrule}m{0.40\textwidth}|}
    \hline
    \rowcolor{tableheader}
        \textbf{Category} & \textbf{Description} & \textbf{Example} \\ \hline
        \textbf{Emotional exploration} & Messages that further inquire about the user's emotional state, mood, and condition. & \tt{``How did you feel after meeting her?''\newline{}``Can you tell me how this situation makes you feel?''} \\ \arrayrulecolor{tablegrayline}\hline
        \textbf{Activity exploration} & Messages exploring the user's experiences or activities. These questions delve into more objective and factual content. & \tt{``What kind of exercise do you do?''\newline{}``Lunch with a friend sounds nice. What did you eat?''} \\ \hline
        \textbf{In-depth follow-up} & Messages that help delve into the root causes or reasons related to mood, feeling, or specific situations. & \tt{``Since when have you felt this overshadowed mood?''\newline{}``If you feel confused, what might be the reason?''} \\ \hline
        \textbf{Future plan exploration} & Messages inquiring about how the user reacted to or plans to respond to the events or emotions they mentioned. & \tt{``That situation must have been challenging. Have you considered any solutions?''\newline{}``What have you done to alleviate the loneliness, even just a bit?''} \\ \arrayrulecolor{black}\hline
    \end{tabular}
    \label{question_type}
\end{table*}

To understand the contents that MindfulDiary generated, we delved deep into the content it generated. 72\% of the AI messages took the form of questions, aiming to elicit responses about users' daily experiences and emotions. We identified and categorized the primary strategies that MindfulDiary employed to assist patients' journaling. There were four strategies employed by the LLM: \textit{Emotional Exploration}, \textit{Activity/Behavior Exploration}, \textit{In-depth Follow-up \& Countermeasures}, and \textit{Future Plan Exploration}. For a comprehensive breakdown of these strategies, along with their descriptions and exemplar questions, refer to \autoref{question_type}.

The average length of participants' responses was 29.42 syllable counts, with a median of 20 ($SD=35.9$). This suggests a left-skewed distribution, where many participants gave shorter responses and a smaller number provided considerably longer answers, causing a high variation. The minimum response length was one character, and the maximum was 559 syllable counts. We further conducted a qualitative analysis of these responses, seeking to identify the themes present in users' interactions with the LLM. This allowed us to understand the scope and topics of the daily records that MindfulDiary collected from the patients.

Participants interacting with MindfulDiary conveyed a range of topics (see \autoref{answer_type}). They described a spectrum of \textit{emotional states}, from negative feelings like exhaustion and anxiety to positive sentiments of pride and joy. \textit{Events and activities} were recounted, offering insights into their daily routines, such as walking during school times or decreased activity post-vacation. They also shared \textit{thoughts and beliefs}, sometimes related to current events, revealing patterns linked to mental health, like feelings of exclusion and loneliness. Regarding \textit{perceived health status}, comments spanned from immediate ailments, such as headaches, to long-term health challenges. Distorted perceptions about their body included content on excessive dieting. Specifically, participants frequently discussed medications, revealing not just their physical reactions but also their perceptions and behaviors toward them. Some expressed concerns over the taste, while others mentioned adverse reactions from intake, like discomfort after swallowing multiple pills at once. Lastly, the realm of \textit{relationships \& interactions} had participants highlighting both the challenges and supports in their interpersonal connections, revealing their significant impact on mental well-being, from conflicts and trust issues to moments of affirmation and encouragement.

\subsection{Patients' Perspectives on MindfulDiary}
Overall, participants viewed MindfulDiary as a space where they could open up and share their stories, feeling a sense of empathy from the system. Participants particularly found the dialogue-driven interactions with MindfulDiary useful. One participant, P15, mentioned, \textit{``If it was just about recording daily activities or emotions like a regular diary, it might have been less engaging, and I could've found it tedious or might not have persisted for long. But this felt like having a conversation with AI, which added an element of fun and kept me engaged in continuous use.''} Such a dialogue-driven journaling process aided participants in maintaining consistent records and helped in forming a habit consistent with our user engagement analysis. P7 stated,\textit{ ``I liked chatting with the AI at first, so I kept using it. The more I used it, the more it became a habit.''}

\begin{table*}[!ht]
    \def\arraystretch{1.2}\setlength{\tabcolsep}{0.25em}
    \centering
    \smaller\sffamily
    \caption{Summary of Participants' Input Messages: Categories include Emotional States, Events \& Activities, Thoughts \& Beliefs, Perceived Health Status, and Relationships \& Interactions with associated descriptions and examples}
    \Description{Table offers a summary of the diverse categories of input messages provided by participants. The table is segmented into three columns: 'Category', 'Description', and 'Example'. 'Emotional States': This category captures the variety of emotions the participants chronicled in their daily interactions. These emotions span a wide array, from negative to depressive feelings. Sample inputs include: “I’m so exhausted, I feel like I’m reaching my limit soon,” and “I’m very worried, scared, and anxious.” 'Events and Activities': This section includes mentions of events, tasks, or activities participants engaged in or observed, such as academic exams or travel. An example message is: “When I attended school, I got some walking in, but after vacation, I don’t have much reason to go out, leading to a decreased activity level.” 'Thoughts and Beliefs': Here, participants' thoughts, values, beliefs, and strong convictions are presented. This category especially encompasses thought patterns linked to mental health like distorted thinking. A representative message is: “I feel like someone is talking behind my back somewhere; they don’t like my actions and seem to exclude me.” 'Perceived Health Status': This category sheds light on participants' physical conditions, health misconceptions, reactions to medications, and their perceptions and behaviors concerning medication intake. Example inputs include: “I’ll starve and exercise to lose weight!” and “I just took my medicine, but it seems to be getting tasteless.” 'Relationships & Interactions': This section brings forward content concerning participants' relationships with others. It covers challenges rooted in interpersonal dynamics and the support and affirmation they receive from acquaintances. An exemplary message is: “I hated seeing my brother being happy. Forcing a cheerful tone also irked me}
    \begin{tabular}{|>{\raggedright\let\newline\\\arraybackslash\hspace{0pt}}m{0.24\textwidth}|m{0.37\textwidth}!{\color{tablegrayline}\vrule}m{0.37\textwidth}|}
    \hline
    \rowcolor{tableheader}
        \textbf{Category} & \textbf{Description} & \textbf{Example} \\ \hline
        \textbf{Emotional states} & The emotions that participants documented in their daily lives encompassed a broad spectrum, ranging from negative and depressive sentiments. & \textit{``I'm so exhausted, I feel like I'm reaching my limit soon.''\newline{}``I'm very worried, scared, and anxious.''} \\ \hline
        \textbf{Events and activities} & The mentions of events, tasks, or activities they participated in or witnessed, such as exam periods or travel. & \textit{``When I attended school, I got some walking in, but after vacation, I don't have much reason to go out, leading to a decreased activity level.''} \\ \hline
        \textbf{Thoughts and beliefs} & The thoughts, values, beliefs, and convictions they usually held. Including characteristic thought patterns related to mental health, such as distorted thinking. & \textit{``I feel like someone is talking behind my back somewhere; they don't like my actions and seem to exclude me.''} \\ \hline
        \textbf{Perceived health status} & The physical state and health concerns, misconceptions about their health, reactions to medications, revealing their perceptions, behaviors related to drug intake.  & \textit{``I'll starve and exercise to lose weight!''\newline{}``I just took my medicine, but it seems to be getting tasteless.''} \\ \hline
        \textbf{Relationships \& interactions} & The contents about relationships with others. Difficulties stemming from interpersonal relationships, as well as the support and affirmation received from those around them. & \textit{``I hated seeing my brother being happy. Forcing a cheerful tone also irked me.''} 
        \arrayrulecolor{black}\\ \hline
    \end{tabular}
    \label{answer_type}
\end{table*}

\subsubsection{Broad Conversational Range: The Versatility in Documenting Diverse Interests}
Our participants appreciated the LLM's flexibility and naturalness in responding to various utterances, topics, and situations. Such broad conversational capabilities of the LLM provided participants with a space where they could document a variety of subjects tailored to individual interests and preferences. In our study, participants interacted with the LLM on diverse topics ranging from games, webcomics, novels, and movies \red{(see \autoref{dial:movie})} to hobbies like Pilates \red{(see \autoref{dial:pilates})}, allowing them to create richer and more personal records. P3 remarked, \textit{``AI systems that I have used in the past could only respond to specific words, but it is amazing how this one can respond to all sorts of things.''} 

\begin{dialogue}[h]
\begin{quotetable}
\quoteuserline{P15}{\red{In the afternoon, I watched movies with my family. We watched two films, Asura and The Witness, which were both quite interesting and thrilling.}}
\quotebotline{\red{Watching movies with family probably made it more enjoyable. 'Asura' and 'The Witness', must have been exciting to watch, since they are both known for their strong stories!}}
\end{quotetable}
\caption{Dialogue snippet from P15}
\label{dial:movie}
\end{dialogue}

\begin{dialogue}[h]
\begin{quotetable}
\quoteuserline{P5}{\red{I've been dancing to exercise, but I don’t think it’s enough.}}
\quotebotline{\red{Dance is a fun and nice form of exercise. But if you feel it's lacking, Why do you think you feel that way?}}
\quoteuserline{P5}{\red{I want to exercise by learning from others, like in Pilates.}}
\quotebotline{\red{You've become interested in Pilates. Can you tell me what makes Pilates interesting to you?}}
\quoteuserline{P5}{\red{I would like to do Pilates, but I can't because there's nowhere to do it.}}
\quotebotline{\red{That must be a difficult situation. What about exercising at home? There are many Pilates videos available.}}
\quoteuserline{P5}{\red{I hadn't thought of following along with videos.}}
\end{quotetable}
\caption{Dialogue snippet from P5}
\label{dial:pilates}
\end{dialogue}

\subsubsection{Expanding Views: Enriching Entries with Varied Perspectives}
Participants also valued the diverse and new perspectives that LLM-generated responses offered, as those helped participants reflect on their struggles, daily events, and emotions from various angles. \red{\autoref{dial:game} shows how the system helps the participant to view the challenges of studying from the perspective of the satisfaction felt in gaming.} This influence helped participants diverge from ruminating on depressive feelings. P12 mentioned, \textit{``Sometimes when you note down emotions, that's the only thought that comes to mind. Beyond that, I don't remember much. Since MindfulDiary uses AI, my thoughts flow more easily, and I like it when it asks about different perspectives or topics.''}. 

\begin{dialogue}[h]
\begin{quotetable}
\quotebotline{\red{Such ordinary days accumulate to form our daily lives. What emotions do you feelwhen you study or play games?}}
\quoteuserline{P2}{\red{I don't like studying, but I'm working hard dreaming about the college I want to go to. Games are, of course, fun and I feel proud when my character grows.}}
\quotebotline{\red{So, studying can be seen as an investment for the future.}}
\quoteuserline{P2}{\red{Right, studying is an investment for the future.}}
\quotebotline{\red{How does it feel when your game character grows? Can you bring that feeling to studying too?}}
\quoteuserline{P2}{\red{The feeling of my game character growing makes me proud, as if I've achieved something. I feel the same pride when I complete my daily study goals.}}
\quotebotline{\red{How about continuing to study with that sense of pride? Could it bring the same sense of achievement?}}
\quoteuserline{P2}{\red{Based on this pride, I plan to focus more on my studies. It won't be easy, but setting clear goals can lead to faster achievements.}}
\end{quotetable}
\caption{Dialogue snippet from P2}
\label{dial:game}
\end{dialogue}

\subsubsection{Probing for Depth: Prompt Questions in Detailed Reflection}
MindfulDiary's question-driven journaling process was also valued by participants as it assisted them with the process of daily reflections and documentation. Compared to their past experiences of journaling, where they had to reflect on their daily life by themselves, participants appreciated that MindfulDiary made the journaling process less daunting. P27 said, \textit{``Because I have to rely solely on my thoughts when I write alone, I sometimes get stuck. But when I was unsure about how to write, the AI helped me. I liked that part.''}. The questions posed by MindfulDiary also guided participants in documenting their daily lives in a more detailed manner by asking their thoughts and feelings about a particular event \red{(See \autoref{dial:stress})}. Such probing allowed for richer, more in-depth entries. P13 shared, \textit{``I used to write diaries on my own and just wrote whatever came to mind. MindfulDiary, however, helped me write in more detail because of the specific questions.''}

\begin{dialogue}[h]
\begin{quotetable}
\quoteuserline{}{\red{[...]}}
\quoteuserline{P13}{\red{I felt so pathetic and wanted to know why I was acting this way.}}
\quotebotline{\red{It must have been really tough to go through such feelings. Does this happen often?}}
\quoteuserline{P13}{\red{It seems to have gotten worse recently.}}
\quotebotline{\red{I see. Do you have any thoughts on why it's getting worse?}}
\quoteuserline{P13}{\red{Lately, I've been mentally exhausted and stressed a lot.}}
\quotebotline{\red{That sounds really difficult. Can you tell me the sources of your stress?}}
\quoteuserline{P13}{\red{Issues with my dorm, relationships with my friends, and worries about the future.}}
\quotebotline{\red{It must be hard with all these concerns overlapping. What's the most pressing issue for you right now?}}
\quoteuserline{P13}{\red{These days, I am very concerned about my relationships with friends.}}
\quotebotline{\red{I understand your worries about friendships. What is your biggest concern?e}}
\quoteuserline{P13}{\red{I feel like people are talking behind my back, disliking my actions, and excluding me from gatherings, which makes me feel very lonely.}}
\quoteuserline{}{\red{[...]}}
\end{quotetable}
\caption{Dialogue snippet from P13}
\label{dial:stress}
\end{dialogue}

\subsubsection{Building Narratives: Structuring Daily Reflections with MindfulDiary}
MindfulDiary's capabilities, such as generating contextualized follow-up questions and summarizing the conversation, made the process easier for participants who struggled to organize daily thoughts and events underpsychotherapy~\cite{donnelly1991cognitive}. In their past experiences, our participants expressed difficulties in journaling because of disjointed thoughts, a lack of clarity in ideas, or inconsistencies in their stories. However, with the support of the LLM in the MindfulDiary, these challenges were addressed, motivating them to record their daily lives persistently. P3 remarked, \textit{`` I often had trouble putting sentences In the past, I would worry about writing the next part. But with this tool, I just tell the story of my day, and it seamlessly continues and wraps it up, presenting a well-structured diary entry. That's its biggest advantage."} \red{(See \autoref{dial:bake})}

\begin{dialogue}[h]
\begin{quotetable}
\quoteuserline{}{\red{[...]}}
\quoteuserline{P3}{\red{The most memorable thing today was my dad eating the cake and food I made yesterday and saying it was delicious.}}
\quotebotline{\red{How was the experience of making the cake? Did you enjoy it?}}
\quoteuserline{P3}{\red{The tough day turned to be enjoyable.}}
\quotebotline{\red{Can you describe more about that feeling of happiness?}}
\quoteuserline{P3}{\red{I felt a bit disappointed and tired while doing it, but when my dad liked it, I felt like I did something great and it made me happy.}}
\quoteuserline{}{\red{[...]}}
\quoteuserline{}{\red{\textbf{[Generated Diary]}\newline{} \texttt{``Today was a day without any special worries. The most memorable thing was my dad eating and complimenting the cake and food I made yesterday. That moment turned a somewhat exhausting and difficult day into a joyful one. My father's praise gave me a great sense of joy and pride. It might be a wrong thought, but I felt really great, like I am an amazing person, which was so nice.''}}}
\end{quotetable}
\caption{Dialogue snippet from P3}
\label{dial:bake}
\end{dialogue}

\subsection{MHPs' Perspectives on MindfulDiary for Clinical Mental Health Settings}
In this section, we describe how MHPs utilized the clinician dashboard and the benefits and drawbacks of the system they reported, drawing on the debriefing interviews with the psychiatrists.

\subsubsection{Utilization of MindfulDiary in Clinic}
During the deployment study, psychiatrists reviewed the journal entries from their patients every morning when they reviewed the medical charts of patients whom they would meet on the day. Depending on the severity and the focal concerns of the patient, psychiatrists spent about 5 to 10 minutes per patient reviewing the MindfulDiary data. After checking trends primarily through PHQ-9 in the clinician dashboard, psychiatrists read summaries about events and documented emotions. If there were spikes or drops in the PHQ-9 or events/emotions, they checked the actual dialogues.

\subsubsection{Percevied Benefits of MindfulDiary for Enhanced Patient Insight and Empathetic Engagement}
All of the psychiatrists emphasized the critical value of an expert interface based on information recorded in the daily lives of patients. Specifically, E3 highlighted MindfulDiary's value in that it consistently aids in recording daily entries, allowing them to utilize more detailed patient data during outpatient visits. \textit{``Patients, with the support of AI, can logically continue their narratives, ensuring more dialogue than a typical (paper-based) diary. This definitely aids me in my consultations.''} (E3). In this section, we further report on how MindfulDiary has been helpful in the clinical practice of psychiatrists.

\ipstart{Enhancing Understanding and Empathy toward Patients}
Psychiatrists indicated that MindfulDiary helped them gain a deeper understanding and empathy about their patients. They perceived that MindfulDiary served as a questioner that could elicit more objective and genuine responses from patients. Psychiatrists appreciated that the LLM was able to pose questions that might be sensitive or burdensome for them to ask, such as patients' negative perceptions of their parents. E4 said:~\textit{``There are times when it's challenging to counter a patient's narrative or offer an opposing perspective. For example, if a patient speaks very negatively about their mother, and we ask, `Didn’t she treat you well when you were younger?', the patient might react aggressively, thinking, `Why is the therapist taking my mother's side?' However, since the LLM is a machine, such concerns are minimized.''}. 

\ipstart{Insights from Everyday Perspectives Outside Clinical Visits}
Psychiatrists valued that MindfulDiary provided them with an understanding of patients' conditions that would be difficult to gain during outpatient visits. For instance, E1 appreciated that MindfulDiary provided them with insights into patients' positive feelings and experiences, which is typically difficult to obtain during clinical consultations. \textit{``Usually, when patients come for a consultation, they talk about bad experiences. Few people come to psychiatry to say, `I've been doing well.' Even if they have good things to say, they usually don't bring them up. But I was happy to see that there were many positive statements in these records, like 'I did that and felt good.' Especially in depression, the presence or absence of positive emotions is crucial. It's a good sign if they show such positive responses.''}. E2 envisioned its potential application to medication management, which is another critical aspect of psychiatric care. He thought these records could be used as a window into understanding how patients react to and perceive medications. For patients undergoing drug therapy, \textit{``If the primary treatment method is pills, but they don’t seem to have an effective response or there's a decline in medication acceptance, I could potentially understand the reasons for it through this diary.''} (E2).

\ipstart{Understanding Patient Progress Through Consistent Record-Keeping}
Feedback from patients highlighted that interactions with MindfulDiary made it easier for patients to maintain a consistent record, as it mitigated the challenges associated with recording. Psychiatrists perceived that having consistent daily data offered them opportunities to observe trends in a patient's condition. E2 said:~\textit{``From our perspective as clinicians, even though we might only see a patient once a month, having access to a record of how they've been throughout the month would allow us to track their progress, which is highly beneficial.''}. In particular, the ability to examine changes not only through quantitative tools like the PHQ-9 but also using a qualitative approach can offer a comprehensive understanding and shed light on the mechanisms influencing a patient's mental health. 

\subsubsection{Perceived Concerns about MindfulDiary}
While MHPs generally appraised the utility of the MindfulDiary positively, they also raised concerns regarding the integration of MindfulDiary into clinical settings.

\ipstart{Significance of Tone and Manner in Patient Data Analysis}
Although patient data summarized and extracted in the expert interface effectively aided in understanding the patient, experts thought that the summarized texts would not convey the patient's tone, pace, and other nuances, which are integral to the Mental Status Examination (MSE) that clinicians utilize. However, MHPs identified the opportunity to perform such analysis from the raw data that patients entered. As the MSE measures objective and quantitative aspects, incorporating such an analysis could make significant improvements in understanding the patient. E1 said, \textit{``In the same way as P14, understanding the tone of this patient may also be possible. That's because we use something called psychiatric MSE, where we observe more than just the patient's appearance, such as tone, pace, and more. Even a short analysis of one's linguistic behavior would be great.''} 

\ipstart{Potential Misuses and Concerns around MindfulDiary}
In our field study, one patient participant perceived the MindfulDiary as a channel to convey their intentions and situations to their psychiatrist. Specifically, the participant, P9, talked to their psychiatrist, \textit{``Have you seen what I wrote?"}, which indicated that the patient was actively attempting to share their current state and situation through MindfulDiary. In spite of the fact that such usage did not seem problematic per se, one psychiatrist raised concerns about the possibility that patients with borderline personality disorders might misuse MindfulDiary as a weapon to manipulate others, such as their providers and parents. \textit{``In some cases, people self-harm out of genuine distress, but others do it to manipulate others, instilling guilt in them so they'll do what they want. There are some patients who write about their distress with sincerity, while there are some who exaggerate their distress in order to get attention.''} For patients exhibiting symptoms of schizophrenia or delusions, there was a concern that MindfulDiary's feature of revisiting past entries could act as a feedback loop, developing and amplifying their delusions. E2 said, \textit{``This diary lets you revisit and organize your past actions. For schizophrenia patients with delusions or unique beliefs, referencing past writings might reinforce their pre-existing delusions. Reaffirming 'Yes, I'm right' can be problematic. The LLM's summaries could exacerbate these delusions if they emphasize distorted content.''}

    \section{Discussion}
In this study, we present MindfulDiary, an LLM-driven journal designed to document the daily experiences of psychiatric patients through naturalistic conversations. Here, we reflect on the opportunities presented by LLM-driven journaling for psychiatric patients and discuss considerations for integrating an LLM-driven patient system into the clinical setting.

\subsection{Guiding Patient Journaling through Conversations Offering Diverse Perspectives}
Our study highlighted the potential of MindfulDiary in clinical settings, mainly where adherence to interventions is important~\cite{musiat}. Core symptoms of depression, such as loss of energy, difficulty in carrying out mental processes, and feelings of apathy, often contribute to lower adherence to a professional’s advice or intervention~\cite{kennedy2022core}. Clinicians who participated in our FGI also highlighted these challenges in motivating patients to utilize the diary writing app. Our findings demonstrated that MindfulDiary helped mitigate these challenges by transforming the conventional journaling process into engaging conversations. Using MindfulDiary, users were able to engage in conversations with the system by answering prompts and questions, which made them feel the journaling process was more accessible and intriguing. This active participation ensures that the users are not overwhelmed by the task and are guided in documenting their feelings and experiences more richly.

Depression often locks patients into negative and rigid thought patterns~\cite{beck1979cognitive}. Such patterns, resistance to change established thought paradigms, can severely limit a patient's ability to perceive issues from multiple angles, leading to a harsh self-judgment~\cite{marazziti2010cognitive}. Our study highlighted that the varied perspectives offered by LLM-driven chatbots like MindfulDiary could help challenge such fixed viewpoints~\cite {10.1145/3532106.3533533}. By prompting users to revisit their initial evaluations or suggest alternative viewpoints, these chatbots could help break the cycle of cognitive rigidity. While our research underscores the promising role of LLM-driven chatbots in assisting psychiatric patients' journaling process, it's essential to note that these are preliminary findings. More work is needed to substantiate these findings in a clinical context.

\subsection{MindfulDiary as a Facilitator for Fostering Patient-Provider Communication}
Studies have suggested that sharing the data captured via chatbots with others, such as health professionals and family members, could further serve as an effective mediator that helps convey more truthful information~\cite{lukoff2018tablechat, 10.1145/3392836}. For instance, patients consistently displayed deep self-disclosure through chatbots, whether or not they intended to share their inputs with health professionals~\cite{10.1145/3392836}. Aligned with prior work on PGHD~\cite{cohen2016integrating, nundy2014using}, MHPs in our study also perceived that MindfulDiary has shed light on patients' daily events, emotions, and thoughts that might have been difficult to gain through regular clinical visits. This data offered MHPs valuable insights into the patient's experiences and context.

\red{Building on these findings, we could expand the potential presented by MindfulDiary in patient-provider communication. In the field of personal health informatics, existing research highlights the role of technology, such as photo journaling, in managing conditions like Irritable Bowel Syndrome. This tool not only empowers patients to record their daily experiences more effectively but also fosters enhanced collaboration between patients and healthcare providers~\cite{10.1145/3123024.3123197, 10.1145/2998181.2998276}. Such tools serve as vital artifacts in negotiating the boundaries of patient-provider interactions (i.e., boundary negotiating artifacts)~\cite{10.1145/2818048.2819926}.} 

\red{This work adds a new dimension to this discussion by showing how LLM-assisted journaling lowers barriers to generating health data in daily life and fosters patient understanding. Specifically, we found that through this system, patients and providers can collaboratively reflect on mental health conditions. In the context of the stage-based model of personal informatics, the patient module in our MindfulDiary helps patients reduce the burden of collecting daily data and supports deeper recording. The expert module's dashboard allows for the combined and transformed processes of diary data, survey data, and quantitative engagement data, supporting MHPs' integration and reflection. Collaborative data generation and utilization with patients can enable care that reflects the patient's values and the characteristics of their daily life. These insights serve as a basis for patient-provider collaboration.}

\red{However, our study findings underscore the importance of careful consideration in the clinical integration of systems like MindfulDiary. While we did not observe patients exaggerating their conditions or needs, this potential issue was raised as a concern by MHPs. They expressed apprehension about the possibility that sharing journal content with MHPs through MindfulDiary might lead some patients to exaggerate their conditions or needs. This concern highlights the need to consider not only the design of chatbots that facilitate patient disclosure behavior but also the complex dynamics between patients and providers in clinical settings. It is crucial to address these dynamics to ensure the effective and safe use of such technologies in mental health care.} The growing prevalence of chatbots in mental health domains emphasizes the need for a holistic approach to their design and implementation. We highlight that engineers and MHPs need to collaborate closely, ensuring that these tools are not only technically sound but also tailored to meet the intricate dynamics of clinical settings~\cite{Thieme2022AIMentalHealth}.

\subsection{Considerations for Integrating LLMs into Clinical Settings} 
In this section, we discuss the consideration for integrating LLMs into clinical mental health settings, drawing insights from the design and evaluation of MindfulDiary.

\ipstart{Aligning Domain Experts' Expectations of LLMs}
Developing and deploying MindfulDiary, we learned that aligning MHPs' expectations with the capabilities and limitations of LLMs involves significant challenges. The capability of generative language models to improve mental health is difficult to measure in comparison with AI models in other medical domains, where objective metrics can determine performance. For instance, in medical imaging, AI can be evaluated based on its accuracy in identifying target diseases from MRI scans, using precise numerical percentages of correct identifications~\cite{shen2017deep}. On the other hand, in the realm of mental health chatbots, gauging success is more nuanced, as it involves subjective interpretations of emotional well-being and psychological improvement, which cannot be easily quantified or compared in the same straightforward manner. This challenge is amplified in mental health, where soft skills like rapport building and emotional observation are important~\cite{10.1007/s40596-016-0627-7}. The use of LLMs in the mental health field is emerging, but little has been said about evaluating or defining the performance of models that are tailored to mental health. Our iterative evaluation process involving MHPs could inform researchers about how to develop and evaluate LLM-mediated mental health technology. When integrating into the clinical setting, this evaluation is also necessary for anticipating who the system would target and for what purpose it would be used. Hence, we advocate that engineers and researchers should carefully consider how to assist domain experts, who may lack AI expertise, in fully and accurately grasping the role and operation of LLM. It is also crucial for researchers and engineers to collaborate closely with these professionals to ensure the technology aligns with therapeutic needs and best practices~\cite {Thieme2022AIMentalHealth}. 

\ipstart{Tailored LLM Evaluation for Clinical Mental Health Domains}
The domain of mental health, which our study addresses, is characterized by the vulnerability of its target user group. The content discussed within this domain is often emotionally charged and sensitive. Therefore, prioritizing user safety becomes even more essential in this domain than in others. Considering the sensitivity of the domain, during our evaluation process, MHPs thoroughly tested the LLM's output by trying out conversations on various sensitive topics in both implicit and explicit ways, drawing upon their clinical experiences. The contents the MHPs input were much more diverse and wide-ranging than what engineers could generate during the development. Additionally, MHPs showed concern that the hallucinations of the LLM could reinforce or expand the delusions of patients with delusional disorders. We highlight that developing evidence-based tests or benchmark sets to anticipate the behavior of the language models in collaboration with MHPs is critical when leveraging LLMs for clinical mental health settings.

\ipstart{Incorporating Perspectives of MHPs in Testing and Monitoring}
Considering the caveats of current LLMs~\cite{npj21}, it is critical to involve MHPs when deploying LLM-driven systems for patients in mental health contexts. While planning the field deployment study of MindfulDiary, we identified specific roles that MHPs could play. In the pre-use phase, MHPs should determine the suitability of users and facilitate the onboarding process with patients. During the mid-use phase, they should closely monitor interactions with the LLM and be prepared to intervene in cases of crises or unexpected use scenarios. Furthermore, they can offer or adjust treatments periodically based on long-term data. Additionally, they should regularly re-evaluate the continued use of the system. While some of these tasks should carefully be designed not to burden MHPs too much, it is important that LLMs do not make autonomous decisions about patients (e.g., diagnosis, prescription, or crisis management) but instead operate under professional oversight.

\ipstart{\changed{Providing Safeguards for Hallucinated LLM Generations}}
\changed{Our clinician dashboard provided various summarized information, such as word cloud, aggregating multiple dialogue entries so that the clinician quickly grasps the gist of the dialogues. Although we underwent intensive testing with the LLM-driven data summarizer, the LLM-driven data processing may still suffer from inaccuracies, biases, and misinterpretation~\cite{rawte2023survey, ji2023survey} of patient sentiments or context, which could adversely affect treatment decisions and patient well-being. 
To mitigate such drawbacks of LLMs in our study, we provided sufficient guidance to MHPs, cautioning them that the LLM-generated information they receive may be error-prone. However, in real-world settings, MHPs might accept the outputs of LLM without much attention. Therefore, when involving LLM-driven data processing, the system should foster careful reviewing of the content based on the expertise of MHPs. For example, future systems could incorporate features like highlighting \textit{in vivo} phrases that were directly mentioned by patients and signify key aspects of their experience and feelings. By contrasting the \textit{in vivo} phrases with the LLM's original text, the system can encourage MHPs to put more scrutiny on the LLM's original interpretation, which may contain errors, and the actual inputs spoken by patients.}

\subsection{Limitations and Future Work}
Our recruitment method could impact the generalizability of our findings, as we recruited the patient participants for our field study from a single university hospital. Although we aimed to recruit patients with diverse types and levels of symptoms, our participants are not representative samples of psychiatric patients. They were young (mostly adolescents) and consulted by a fixed number of psychiatrists. While this work is just a first step toward designing an LLM-driven journaling app for psychiatric patients, further investigation is necessary with subjects from various backgrounds. To implement our pipeline, we used OpenAI's GPT API, which provided the most capable LLM at the time of our study and was accessible via commercial API. As GPT models are continually updated, later models may not yield the same conversational behavior. To generalize the performance of our conversational pipeline design, future work is needed to compare multiple versions of MindfulDiary with different underlying LLMs.

    \section{Conclusion}
In this paper, we designed MindfulDiary to assist psychiatric patients undergoing outpatient treatment with journaling in their daily lives. Keeping the clinical mental health setting in mind, our system was developed in collaboration with MHPs, from the initial concept building to the design of LLM's conversation flow and evaluation. MindfulDiary leverages a stage-based LLM-driven chatbot, enabling patients to interact through prompt questions and answers, while complying with guidelines based on MHPs and literature. We conducted a field deployment study with 28 patients over 4 weeks. We found that the versatility, narrative-building capability, and diverse perspectives provided by MindfulDiary assisted patients in consistently enriching their daily records. The enriched records from MindfulDiary provided psychiatrists with deeper insights, enhancing their understanding and empathy toward their patients. We hope that this research provides a case study and insight into the development of an LLM-driven chatbot for mental health that is clinically relevant and reflects the needs and experiences of MHPs.
\begin{acks}
We thank our study participants for their time and efforts. We also thank Eunkyung Jo and Yubin Choi for providing feedback on the early draft of this paper. This work was supported through a research internship at NAVER AI Lab, and in part supported by a grant of the Korea Health Technology R\&D Project through the Korea
Health Industry Development Institute (KHIDI), funded by the Ministry of Health \& Welfare, Republic of Korea (grant number : HI22C1962) and by the National Research Foundation of Korea grant funded by the Ministry of Science and ICT (No. 2022R1G1A1009100).
\end{acks}

\bibliographystyle{ACM-Reference-Format}
\bibliography{sample-base}

\appendix

\end{document}